\documentclass[12pt,doublespacing]{article}

\usepackage{fullpage}
\usepackage{setspace}
\usepackage{authblk}

\usepackage{cite}
\usepackage{hyperref} 

\usepackage{graphicx}

\usepackage{amsmath}
\usepackage{amssymb}

\usepackage{float}
\usepackage{subfig}
\usepackage{makecell}
\usepackage{rotating}

\usepackage{enumitem}

\begin{document}

\title{Multiband Hybrid Metasurface for Enhanced Second-Harmonic Generation via Coupled Gap Surface Plasmon Modes}

\author[1]{Partha Mondal}
\author[2]{Omar Alkhazragi}
\author[1]{Boon S. Ooi}
\author[1]{Hakan Bagci\vspace{0.5cm}}

\affil[1]{Electrical and Computer Engineering (ECE) Program 
\authorcr Computer, Electrical and Mathematical Sciences and Engineering (CEMSE) Division
\authorcr King Abdullah University of Science and Technology (KAUST)
\authorcr Thuwal 23955, Saudi Arabia
\authorcr e-mail: partha.mondal@kaust.edu.sa \vspace{1cm}}

\affil[2]{Department of Electrical Engineering
\authorcr King Fahd University of Petroleum and Minerals (KFUPM), Dhahran 31261, Saudi Arabia}

\date{}
\maketitle
\newpage

\begin{abstract}
A multiband hybrid metasurface supporting multiple gap-surface plasmon (GSP) and localized surface plasmon (LSP) modes is presented. The structure adopts a metal–dielectric–metal configuration consisting of an aluminum bottom layer, a silicon dioxide spacer, and a bar--disc hybrid resonator patterned in the top aluminum layer. Optimized geometrical parameters yield four distinct resonances across the near-infrared and telecommunication bands, arising from the interplay between GSP modes and LSP excitations. The reflectance spectra are systematically analyzed as functions of geometric parameters and polarization, demonstrating tunable multiband operation. Experimental measurements of the fabricated metasurface show good agreement with numerical predictions. Furthermore, the second-harmonic generation (SHG) response is numerically investigated, revealing enhanced SH emission at the resonance wavelengths due to strong electromagnetic field confinement within the metal–dielectric–metal cavity. The proposed metasurface provides a compact platform for multiband and multifunctional nanophotonic applications. 
\end{abstract}
\newpage

\section{Introduction}
In recent years, nanostructured plasmonic metasurfaces have attracted considerable attention due to their ability to manipulate electromagnetic fields at subwavelength scales~\cite{meinzer2014plasmonic,yang2025plasmonic}. 
By engineering meta-atoms with tailored geometries, these platforms control wavefront, amplitude, and polarization through physical mechanisms, including surface plasmon polaritons (SPPs), localized surface plasmons (LSPs), and cavity resonances~\cite{lameirinhas2022new, bin2021ultra}. Among these, gap surface plasmons (GSPs), which originate from electromagnetic coupling between closely spaced metallic layers,  have emerged as a particularly powerful mechanism~\cite{lei2012revealing, yeshchenko2015surface}. GSP-based metasurfaces, typically realized in metal--dielectric--metal configurations and compatible with a single-step lithographic fabrication, enable precise manipulation of the phase, amplitude, and polarization of reflected light~\cite{ding2019review,wang2025double,liu2025dual}. As a result, they have been widely employed in applications, such as beam steering~\cite{meng2020optical,ding2020gap, deng2021functional,ding2018bifunctional}, metalens design~\cite{tang2020high,lu2018broadband}, hologram generation~\cite{zheng2015metasurface,chen2014high,im2023broadband}, structural color printing~\cite{deshpande2019plasmonic,huang2020polarization}, polarization conversion~\cite{ding2020gap,heidari2024broadband}, broadband perfect absorption~\cite{nielsen2012efficient,nejat2020sensing,nourbakhsh2020ultra,guo2017electrically}, and optical sensing~\cite{chou2021significantly,nejat2020sensing}.

Beyond linear light–matter interactions, the strong subwavelength field confinement supported by plasmonic nanostructures enables enhanced nonlinear optical responses. SPPs provide an effective mechanism for wavelength conversion processes, such as second-harmonic generation (SHG)~\cite{sugita2023surface,zhang2024plasmon,guo2020graphene,sun2024230} and third-harmonic generation (THG)~\cite{spear2023surface,gour2022enhancement}. In parallel, extensive studies have demonstrated nonlinear responses arising from LSP resonances~\cite{chu2022second,nezami2015localized,stolt2022multiply}. More recently, GSP modes have emerged as a particularly efficient platform for enhancing nonlinear effects~\cite{dass2019gap,lee2014giant,nezami2016gap,zeng2018enhanced}. The strong electromagnetic confinement within the dielectric gap, combined with the intrinsic surface nonlinearities of metals, increases the efficiency of harmonic generation processes. Consequently, GSP-based metasurfaces provide a promising platform for enhancing nonlinear optical phenomena~\cite{rahimi2018nonlinear}.

Metasurfaces supporting multiple resonances within a single structural platform are highly desirable for applications requiring control of electromagnetic response over a broad frequency range~\cite{zhao2018design, duportal2024multi, guo2024multiband, fu2025multi, wang2024dynamically,nagini2022wideband}. A single meta-atom supporting multiple resonant modes enables diverse functionalities, including spectroscopic detection~\cite{tali2019multiresonant, islam2020tunable}, multicolor light manipulation~\cite{dayal2017high}, and nonlinear frequency conversion~\cite{liu2016polarization}. In particular, multiresonant metasurfaces provide an efficient platform for realizing broadband nonlinear optical responses~\cite{huttunen2019efficient,aouani2012multiresonant}. Furthermore, in doubly resonant configurations, distinct plasmonic modes are engineered to coincide with both the fundamental and harmonic wavelengths, thereby substantially increasing the nonlinear conversion efficiency~\cite{liu2016polarization,thyagarajan2012enhanced,yang2017enhancement}. One approach to achieving multiband operation is to integrate simple or complex meta-atoms, each designed to support a distinct resonance~\cite {wang2016visible,zhu2024compound,aieta2015multiwavelength}. 

However, despite these advantages, implementing multiresonant behavior in GSP-based metasurfaces is intrinsically challenging due to the dispersion-limited response of their constituent meta-atoms. Because the reflection phase strongly depends on the geometrical dimensions of the resonators, the achievable phase gradient rapidly degrades at wavelengths detuned from the resonance condition, restricting broadband operation. To date, metal--dielectric--metal nanostructures supporting GSP modes have primarily demonstrated single- or dual-band absorption~\cite{tang2018polarization,wang2025wide}. For example, a dual-band metasurface that supports first- and third-order GSP modes in a metal--dielectric--metal configuration has been reported~\cite{deshpande2019dual}. An alternative strategy incorporates detuned resonators to extend the accessible phase range and partially alleviate bandwidth limitations. Although this approach improves phase coverage and operational bandwidth compared to single-resonator designs~\cite{damgaard2020demonstration}, achieving a compact nanophotonic platform with robust multiband and multifunctional performance remains challenging. 

This work presents a hybrid GSP-based metasurface design that overcomes bandwidth limitations while enabling multiresonant operation within a single meta-atom. The proposed architecture integrates a metallic bar and a metallic disc into the top layer of a metal--dielectric--metal configuration, forming a coupled hybrid resonator. Numerical analysis shows that the metasurface supports four distinct resonances across the near-infrared and telecommunication bands. These resonances arise from the interplay between LSP and GSP modes supported by the hybrid resonator. 

The novelty of the present work lies in the hybrid integration of GSP and LSP modes within a compact meta-atom, enabling four distinct resonances spanning the near-infrared and telecommunication bands. Unlike conventional approaches relying on independent resonators, the proposed bar–disc hybrid configuration exploits strong electromagnetic coupling between LSP and GSP modes, resulting in enhanced field confinement. Furthermore, the metasurface exhibits enhanced SHG at multiple resonances, enabled by strong field localization within the metal-dielectric-metal cavity. The proposed design is experimentally validated, with good agreement between simulation and experiment confirmed across all resonant modes.

The dependence of the resonant behavior on structural parameters and polarization is systematically investigated. The proposed multiband LSP–GSP metasurface provides a compact platform for multifunctional applications requiring extended frequency coverage.

\section{Metasurface Design and Simulation}\label{metasurface_simulation}

\begin{figure}[ht!]
\centering
\subfloat[]{\includegraphics[width=0.40\columnwidth]{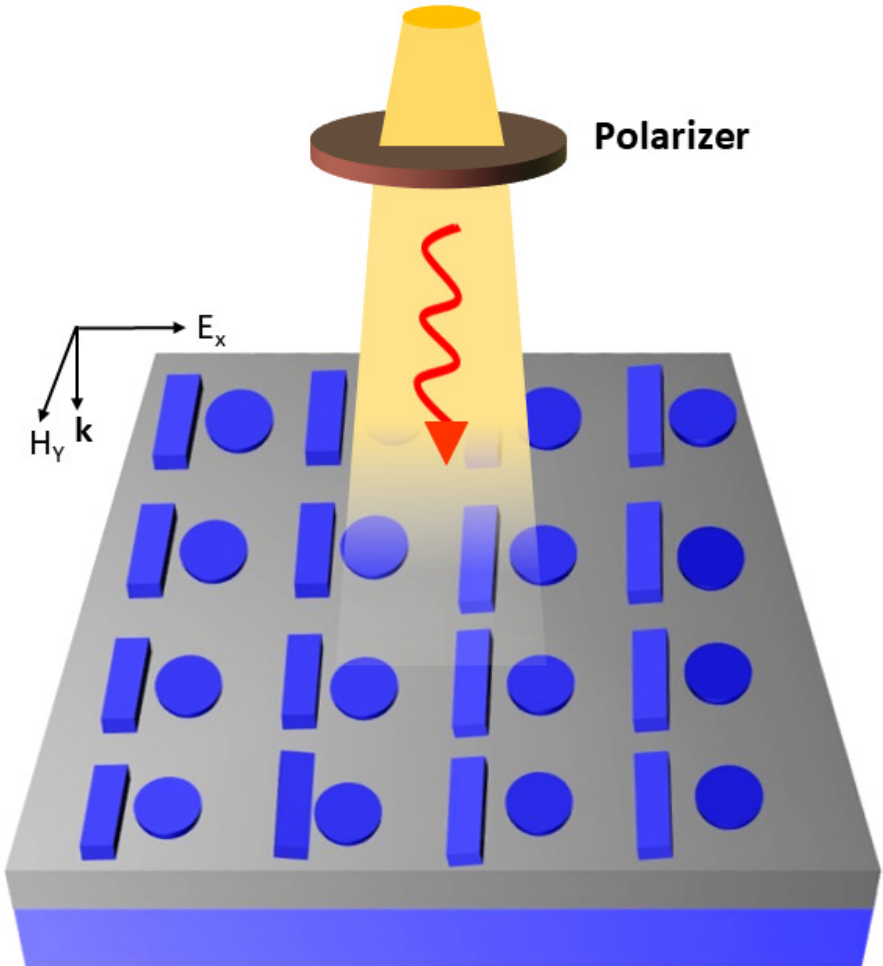}}\\
\subfloat[]{\includegraphics[width=0.50\columnwidth]{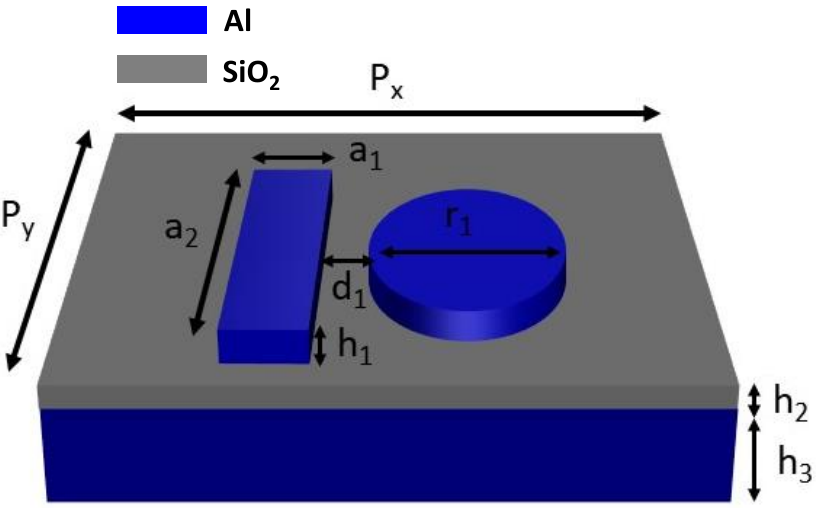}}
\caption{(a) Three-dimensional schematic of the proposed hybrid metasurface. The top $\mathrm{Al}$ layer consists of a bar--disc hybrid resonator, separated from an optically thick continuous bottom $\mathrm{Al}$ layer by a $\mathrm{SiO_{2}}$ dielectric spacer, forming a metal–dielectric–metal configuration that supports GSP modes. (b) Unit-cell geometry showing the structural parameters: bar width $a_1$, bar length $a_2$, disc radius $r_1$, bar-disc separation $d_1$, layer thicknesses $h_1$, $h_2$, and $h_3$, and lattice periods $P_x=P_y$.}
\label{design_structure}
\end{figure}

Fig.~\ref{design_structure}(a) presents a three-dimensional schematic of the proposed hybrid metasurface, while Fig.~\ref{design_structure}(b) shows the corresponding unit-cell geometry with the structural parameters. The metasurface consists of a periodic array in which each unit cell contains a bar--disc hybrid plasmonic resonator made of aluminum ($\mathrm{Al}$). These resonators are positioned above a continuous $\mathrm{Al}$ bottom layer that is sufficiently thick to suppress optical transmission. A thin $\mathrm{SiO_{2}}$ dielectric spacer is inserted between the two metallic layers, forming a metal--dielectric--metal configuration. The metal--dielectric--metal configuration enables strong electromagnetic confinement within the spacer layer, supporting GSP modes. 

The electromagnetic response of the metasurface is numerically investigated using the finite-difference time-domain (FDTD) method~\cite{fdtd2005, vial2007description}. To model the periodic array, periodic boundary conditions are applied along the $x$- and $y$-directions, while perfectly matched layers are used along the $z$-direction. The structure is excited by a normally incident plane wave propagating along the $z$-direction and polarized along the $x$-direction. The optimized structural parameters used to achieve multiband operation are: top $\mathrm{Al}$ thickness $h_{1} = 30\, \mathrm{nm}$, $\mathrm{SiO_{2}}$ dielectric spacer thickness $h_{2} = 70\, \mathrm{nm}$, bottom $\mathrm{Al}$ thickness $h_{3} = 150 \, \mathrm{nm}$, bar width $a_{1} = 180\, \mathrm{nm}$, bar length $a_{2} = 485\, \mathrm{nm}$, bar--disc separation distance $d_{1} = 50\, \mathrm{nm}$, and unit-cell periods $P_{x}$ = $P_{y} = 900\, \mathrm{nm}$. 

The optical response of $\mathrm{Al}$ in the near-infrared region, particularly around $800$--$850\,\mathrm{nm}$ ($\sim 1.5\,\mathrm{eV}$), is strongly influenced by interband transitions. A simple Drude model, which accounts only for free-electron contributions, is therefore insufficient to accurately describe the associated optical losses. To address this, all FDTD simulations employ the frequency-dependent complex permittivity of $\mathrm{Al}$ from the experimentally measured optical constants reported in~\cite{palik1998handbook}. These optical constants inherently incorporate interband transition effects and provide a physically realistic representation of the material response over the spectral range of interest. 

When linearly polarized light is normally incident on the metal--dielectric-metal structure, LSP oscillations are excited at the edges of the top metallic resonators. These oscillations induce strong charge accumulation at the metal-dielectric interfaces and provide lateral confinement for GSP modes supported within the dielectric spacer. As a result, standing-wave patterns form beneath the resonators due to reflections at their terminations, leading to the excitation of laterally confined GSP cavity modes. The resonance condition of these GSP modes can be approximately described using a Fabry-Pérot cavity model~\cite{deshpande2019dual}:
 \begin{equation}
   \beta_{\mathrm{gsp}} w  + \phi= p \pi
   \label{eqn:FP}
 \end{equation}
 where $\beta_{\mathrm{gsp}}$ is the in-plane propagation constant of the GSP mode in the dielectric spacer, $w$ represents the effective lateral cavity length defined by the resonator dimensions, $\phi$ is the phase shift upon reflection at the resonator edges, and $p$ is the integer mode order.

\begin{figure}[ht!]
  \centering
   \subfloat[]{\includegraphics[width=0.65\columnwidth]{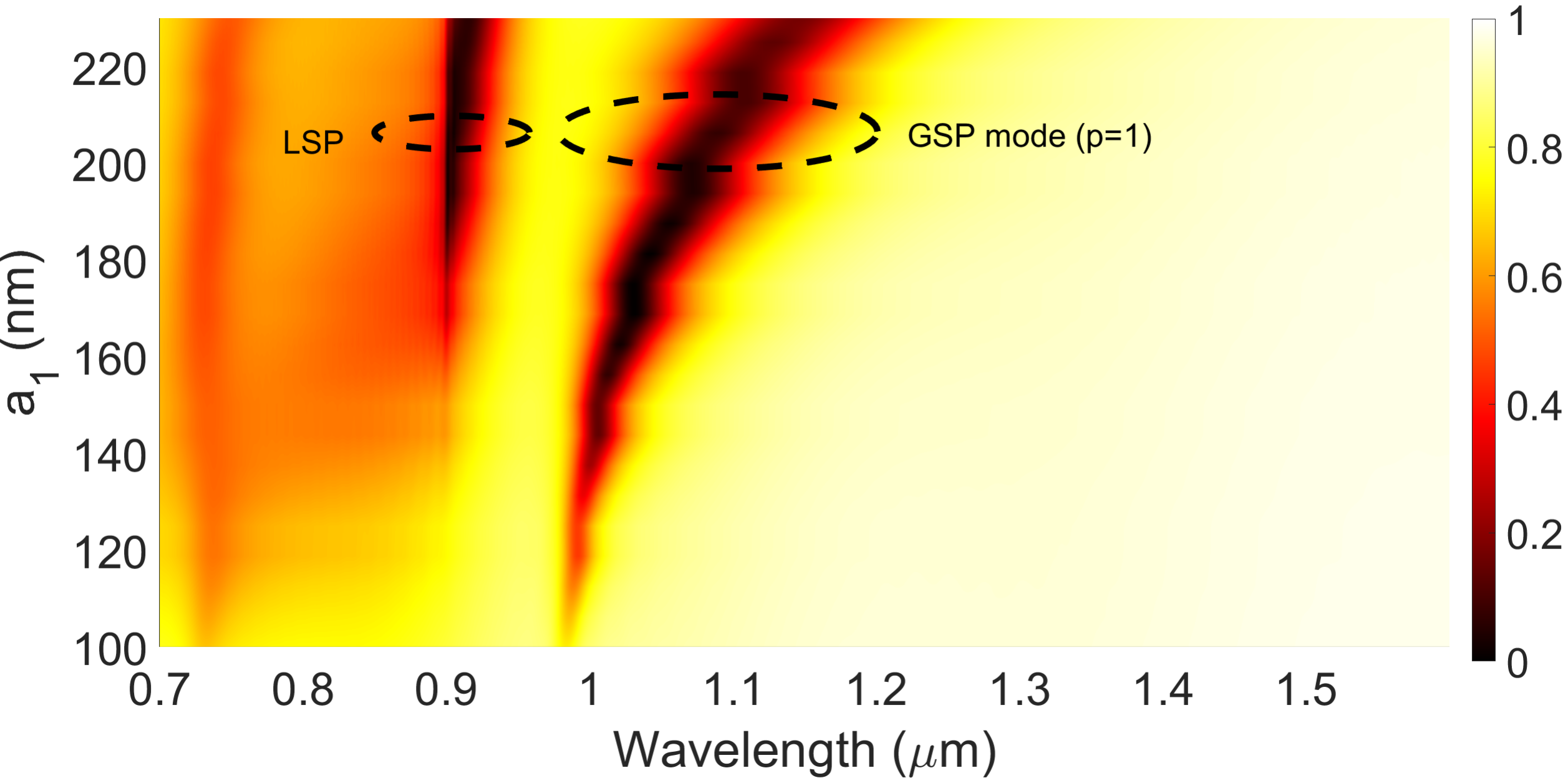}}\\
   \subfloat[]{\includegraphics[width=0.65\columnwidth]{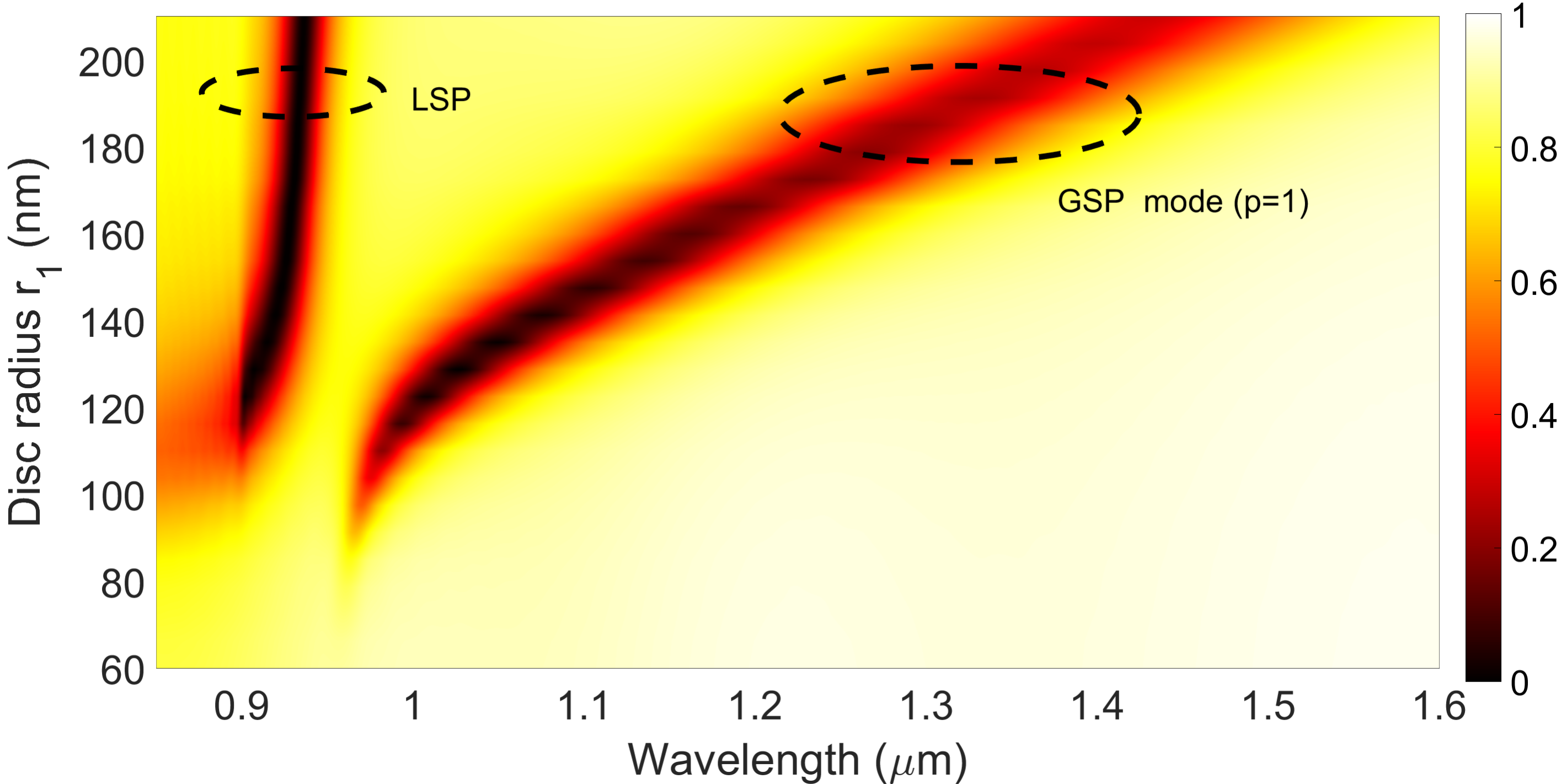}}\\
   \subfloat[]{\includegraphics[width=0.65\columnwidth]{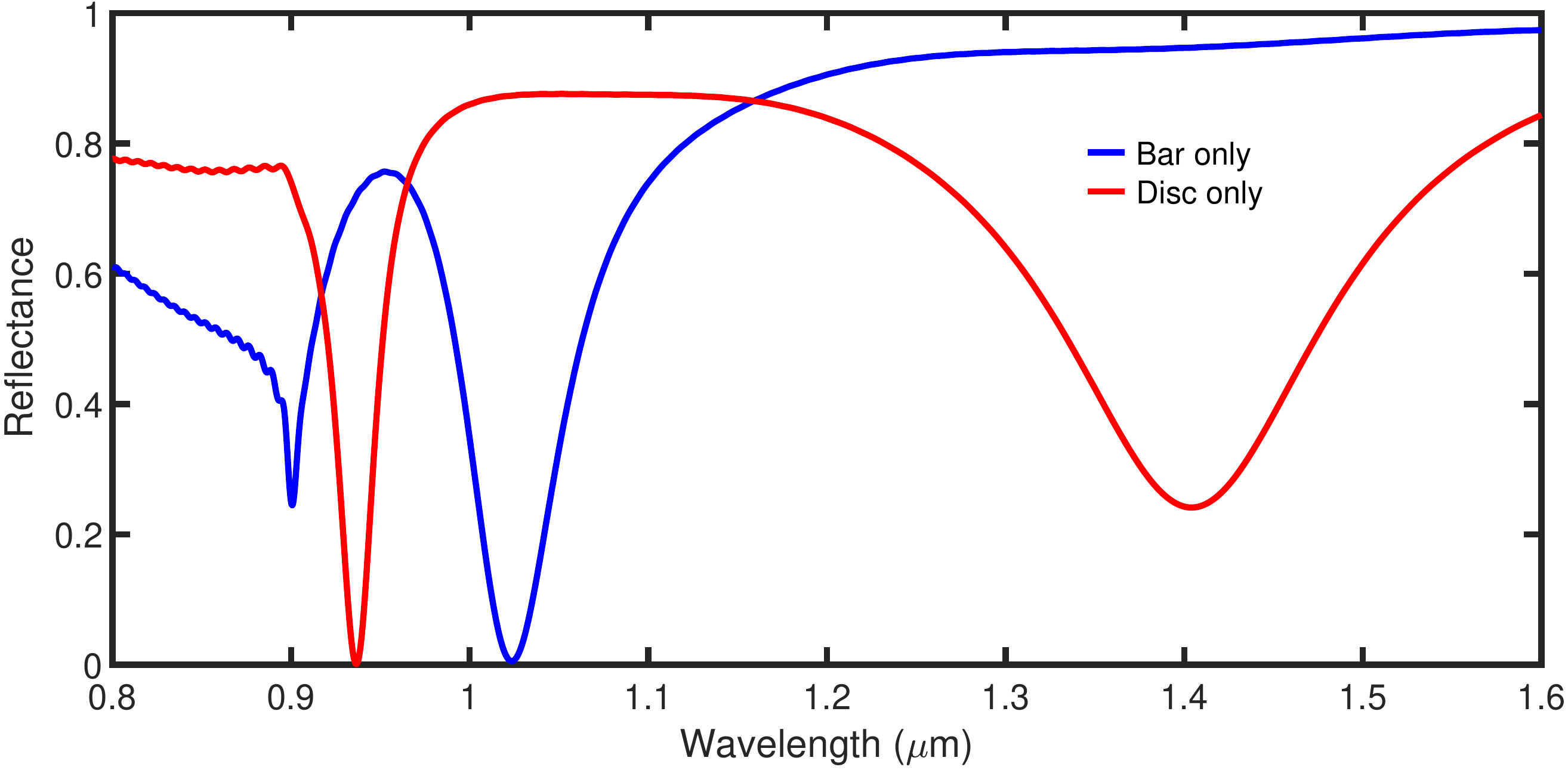}}
  \caption{Simulated reflectance spectra of the isolated resonators. (a) Bar-only structure for varying bar width $a_1$. (b) Disc-only structure for varying disc radius $r_1$. (c) Reflectance spectra of the isolated bar and disc resonators with optimized geometrical parameters.}
  \label{bar_disc_only_colorplot}
\end{figure}

To elucidate the contribution of individual resonators to the overall optical response, electromagnetic simulations of isolated bar and disc resonators are performed. This analysis provides insight into the resonance characteristics of each constituent resonator. Fig.~\ref{bar_disc_only_colorplot}(a) presents the reflectance spectra of the bar resonator as a function of its width $a_{1}$, with its length $a_{2}$ fixed at $500$ $\mathrm{nm}$. Increasing  $a_{1}$ produces a pronounced redshift of the GSP resonance, consistent with the increase in the effective lateral cavity length governing the Fabry–Pérot condition~\eqref{eqn:FP}. In contrast, the resonance associated with LSP excitation exhibits only weak spectral variation, indicating that it is primarily determined by the intrinsic dipolar response of the bar. Similarly, the reflectance spectra of the disc, shown in Fig.~\ref{bar_disc_only_colorplot}(b), demonstrate that increasing the disc radius $r_1$ leads to a redshift of the GSP mode, while the LSP resonance remains comparatively less sensitive to geometrical variation. 

Fig.~\ref{bar_disc_only_colorplot}(c) displays the reflectance spectra of the individual bar and disc resonators with optimized geometrical parameters, revealing multiple distinct resonances spanning the near-infrared to telecommunication range. The specific modal nature of these resonances is clarified through the field-distribution analysis presented in the following section.

\begin{figure}[ht!]
\centering
\subfloat[]{\includegraphics[width=0.48\columnwidth]{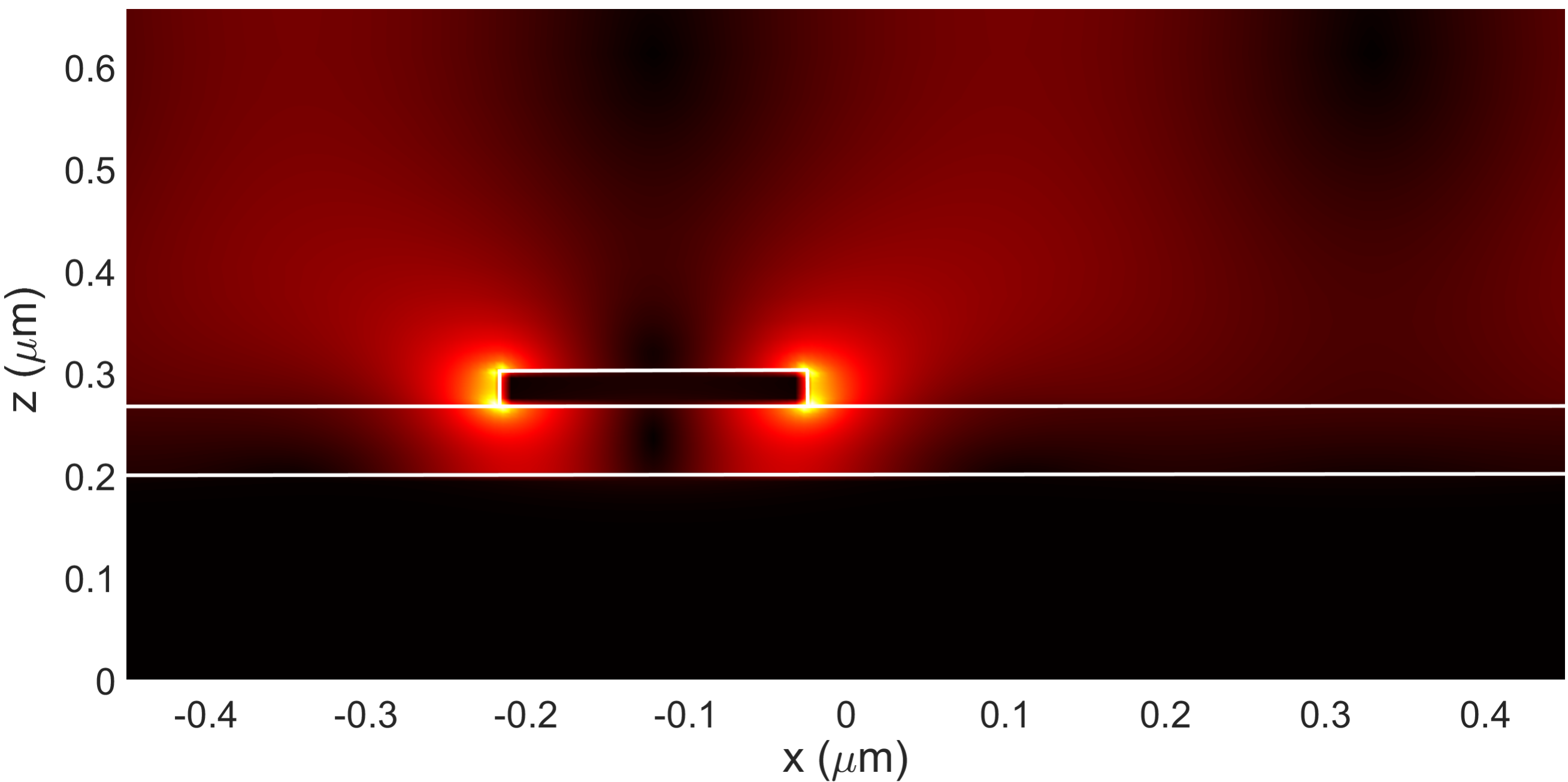}}\hspace{4pt}
\subfloat[]{\includegraphics[width=0.48\columnwidth]{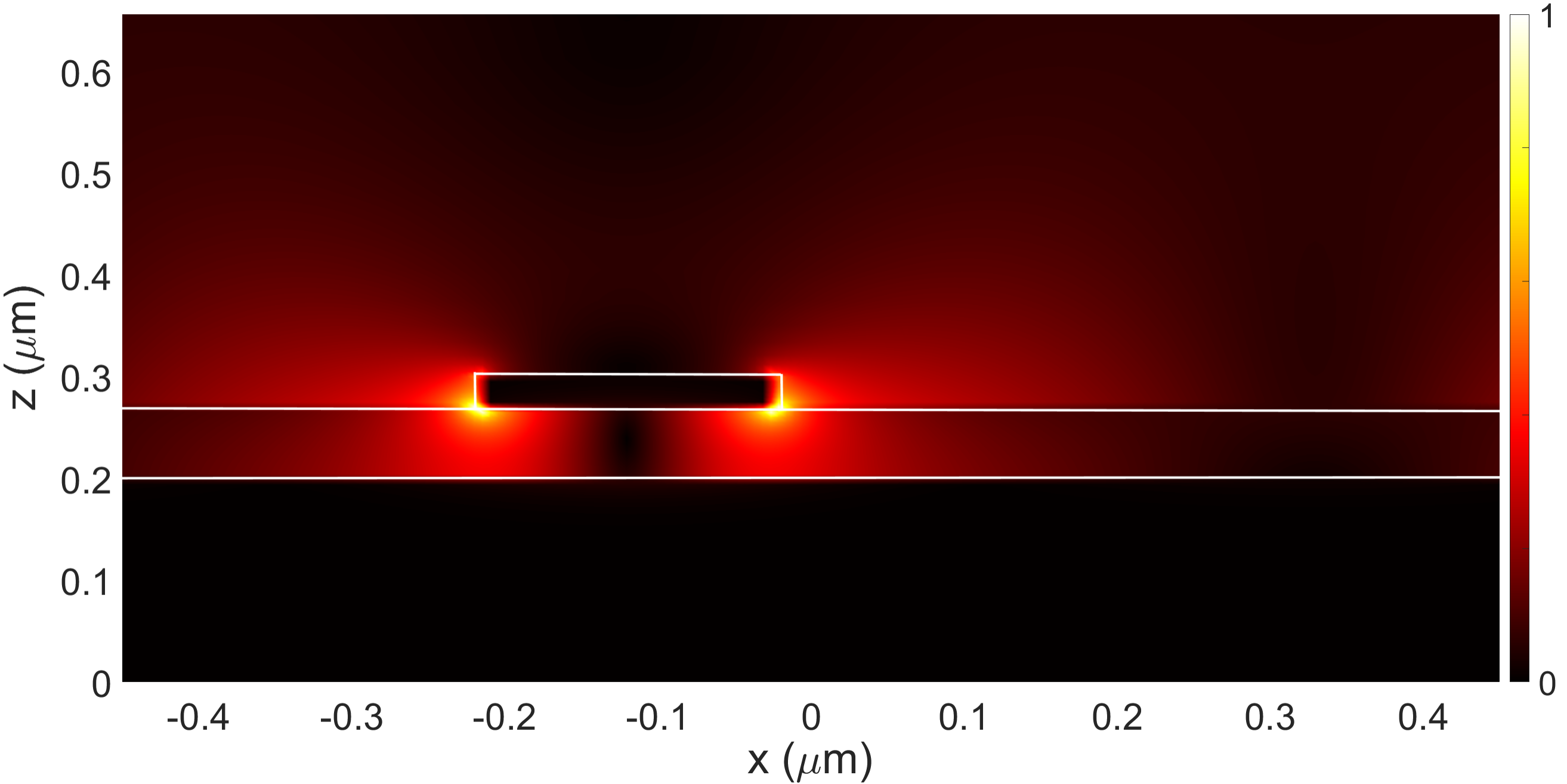}}\\
\subfloat[]{\includegraphics[width=0.48\columnwidth]{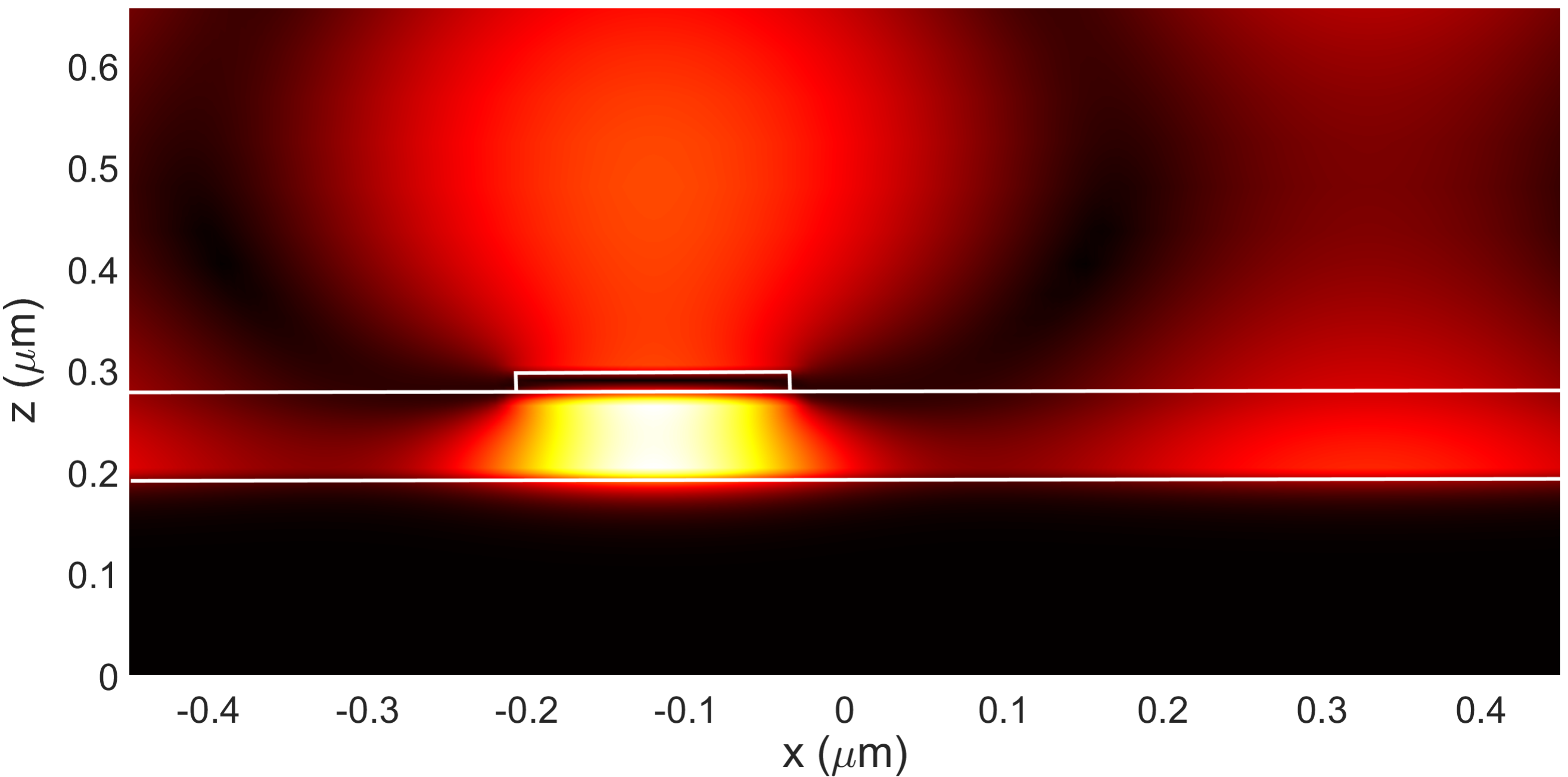}}\hspace{4pt}
\subfloat[]{\includegraphics[width=0.48\columnwidth]{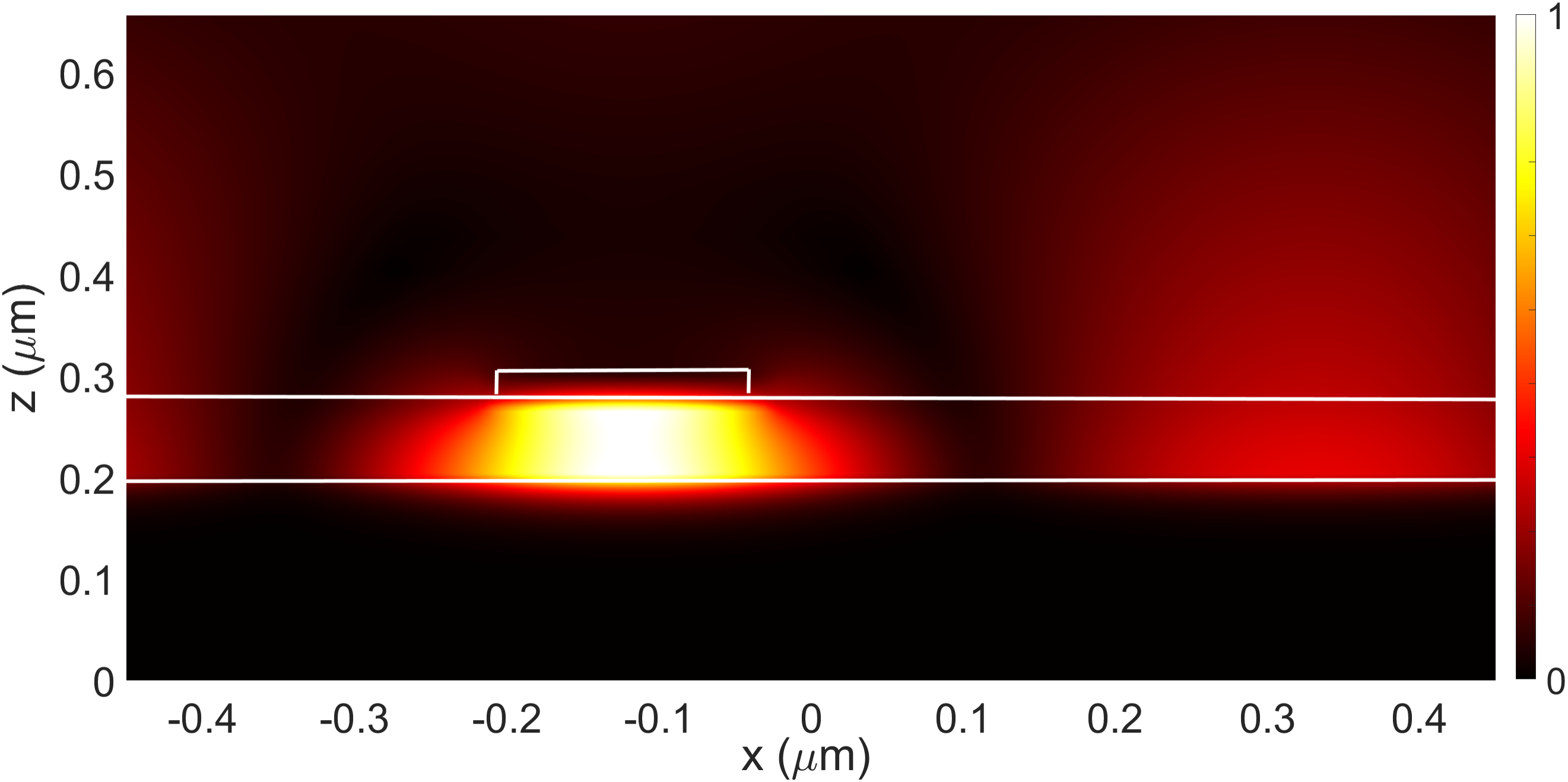}}
\caption{Electromagnetic field distributions of the isolated bar resonator in the $x z$-plane at two resonant wavelengths. (a-b) Normalized electric-field distributions. (c-d) Corresponding normalized magnetic-field distributions.}
\label{Bar_electric_magnetic_field_combined}
\end{figure}

To verify the physical origin of the observed resonances, the corresponding electromagnetic field distributions are examined at their respective resonance wavelengths. Fig.~\ref{Bar_electric_magnetic_field_combined} shows the normalized electric- and magnetic-field distributions in the $xz$-plane for two resonances of the isolated bar resonator. The electric-field distribution in Fig.~\ref{Bar_electric_magnetic_field_combined}(a) exhibits strong localization near the bar edges, while Fig.~\ref{Bar_electric_magnetic_field_combined}(c) shows relatively weak magnetic-field confinement within the dielectric spacer, with significant radiation leakage above the resonator. This behavior indicates that the resonance is primarily associated with LSP excitation of the resonator. In contrast, Fig.~\ref{Bar_electric_magnetic_field_combined}(b) and~\ref{Bar_electric_magnetic_field_combined}(d) exhibits pronounced electric- and magnetic-field confinement within the dielectric spacer beneath the bar, respectively. The magnetic field forms a distinct standing-wave pattern characterized by a single antinode, confirming the excitation of a first-order GSP mode ($p=1$).

\begin{figure}[t!]
\centering
\subfloat[]{\includegraphics[width=0.48\columnwidth]{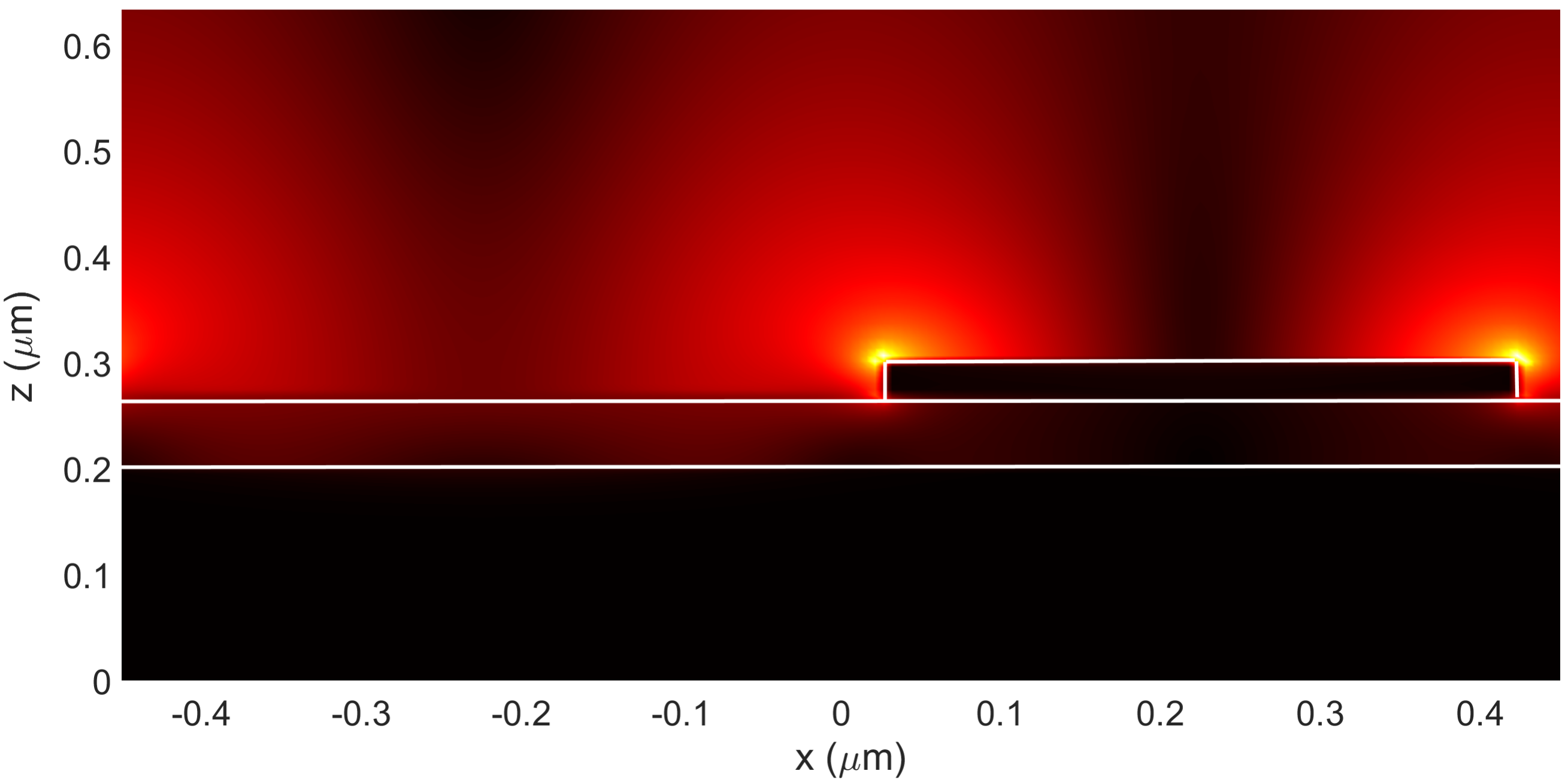}}\hspace{4pt}
\subfloat[]{\includegraphics[width=0.48\columnwidth]{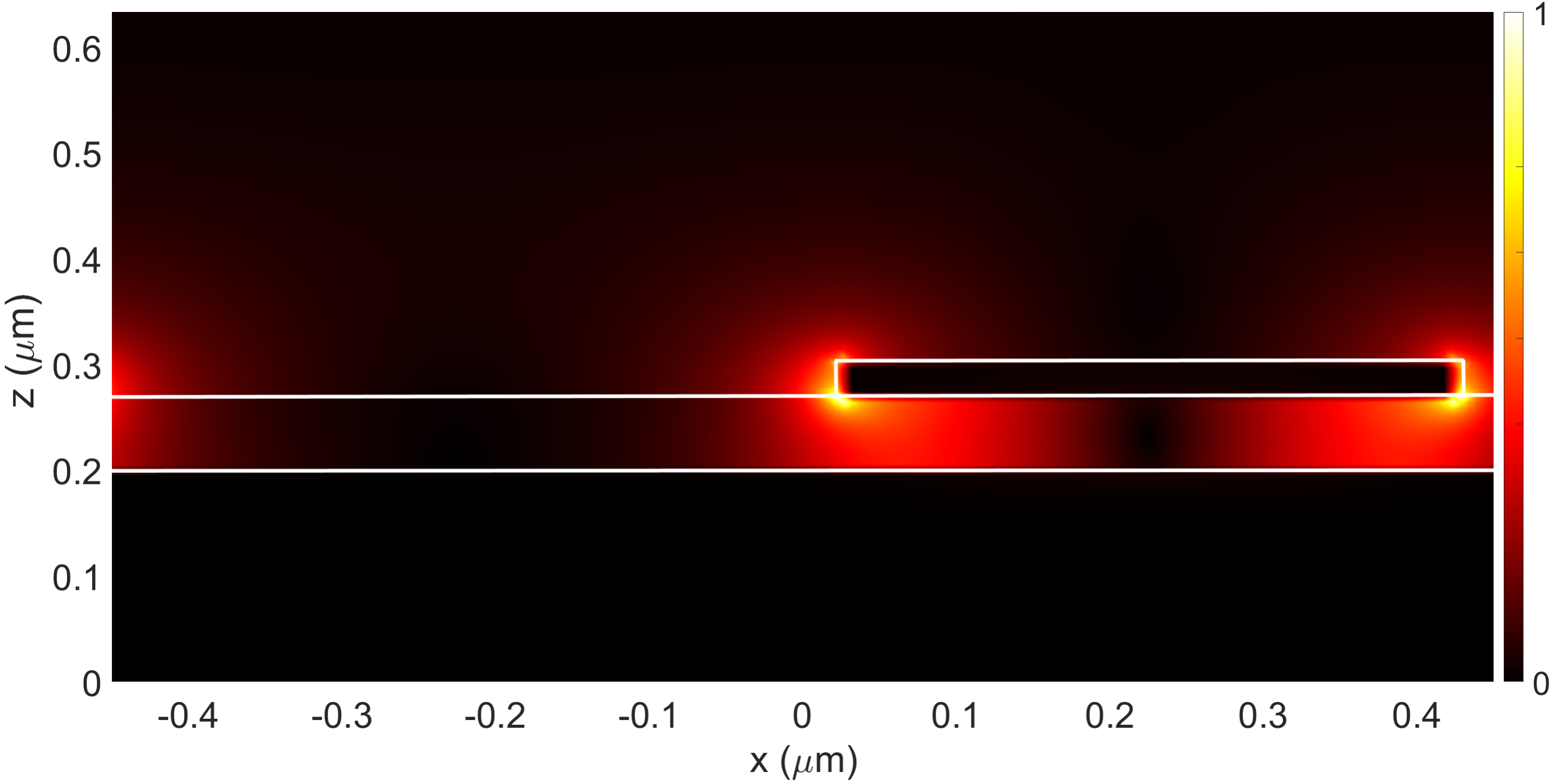}}\\
\subfloat[]{\includegraphics[width=0.48\columnwidth]{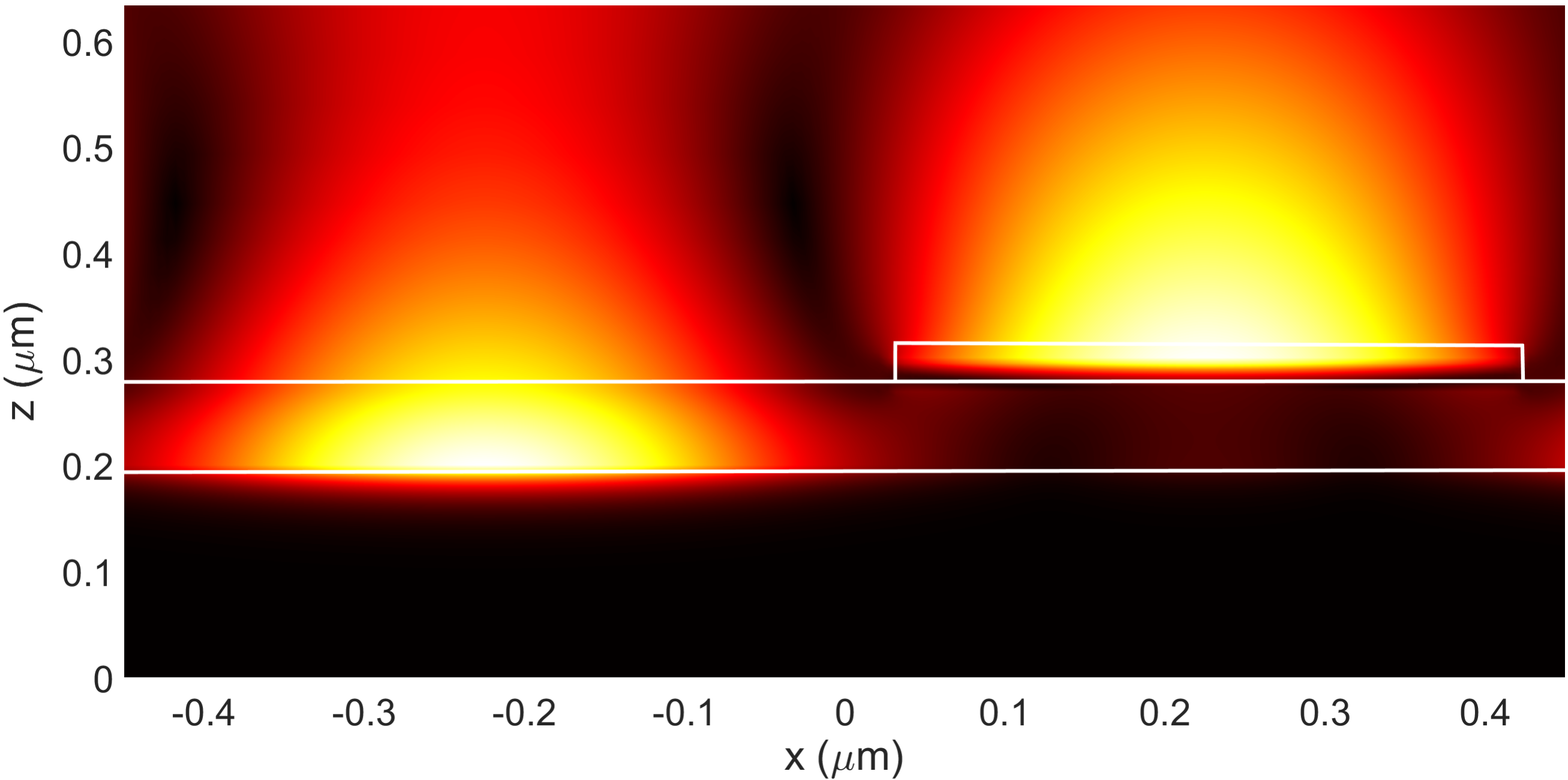}}\hspace{4pt}
\subfloat[]{\includegraphics[width=0.48\columnwidth]{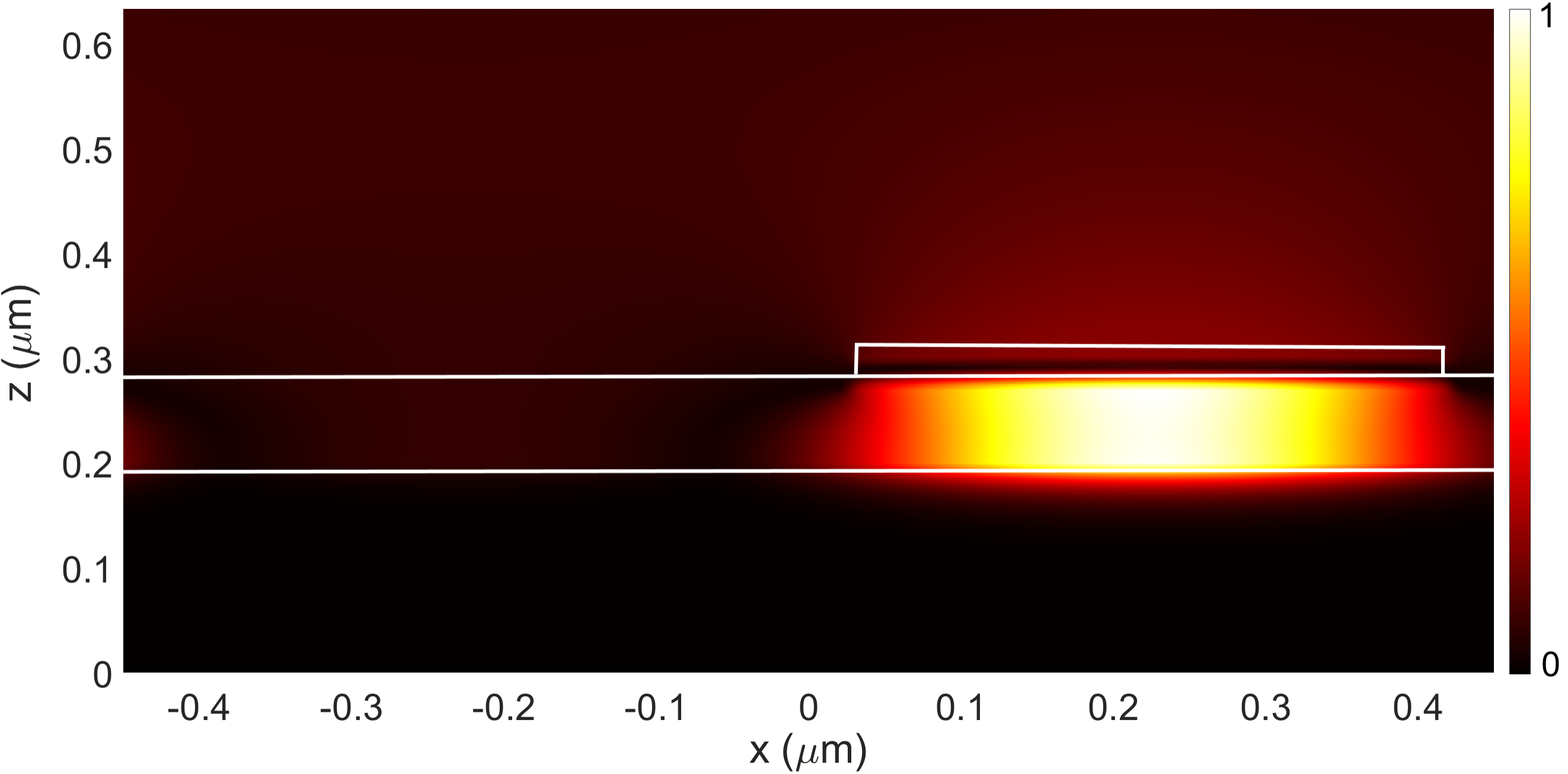}}
\caption{Electromagnetic field distributions of the isolated disc resonator in the $x z$-plane at two resonant wavelengths. (a-b) Normalized electric-field distributions. (c-d) Corresponding normalized magnetic-field distributions.}
\label{Disc_electric_magnetic_field_combined}
\end{figure}

Similarly, Fig.~\ref{Disc_electric_magnetic_field_combined} presents the normalized electric- and magnetic-field distributions for the isolated disc resonator. The field profiles in Fig.~\ref{Disc_electric_magnetic_field_combined}(a) and~\ref{Disc_electric_magnetic_field_combined}(c) exhibit strong electric-field localization near the disc edges and weak magnetic confinement under the disc within the spacer, indicating excitation of an LSP mode. In contrast, Fig.~\ref{Disc_electric_magnetic_field_combined}(b) and~\ref{Disc_electric_magnetic_field_combined}(d) shows pronounced electric- and magnetic-field confinement within the dielectric spacer, with a clear standing-wave pattern characterized by a single magnetic antinode, confirming the excitation of the first-order GSP mode ($p=1$).

\begin{figure}[t!]
  \centering
   \includegraphics[width=0.85\columnwidth]{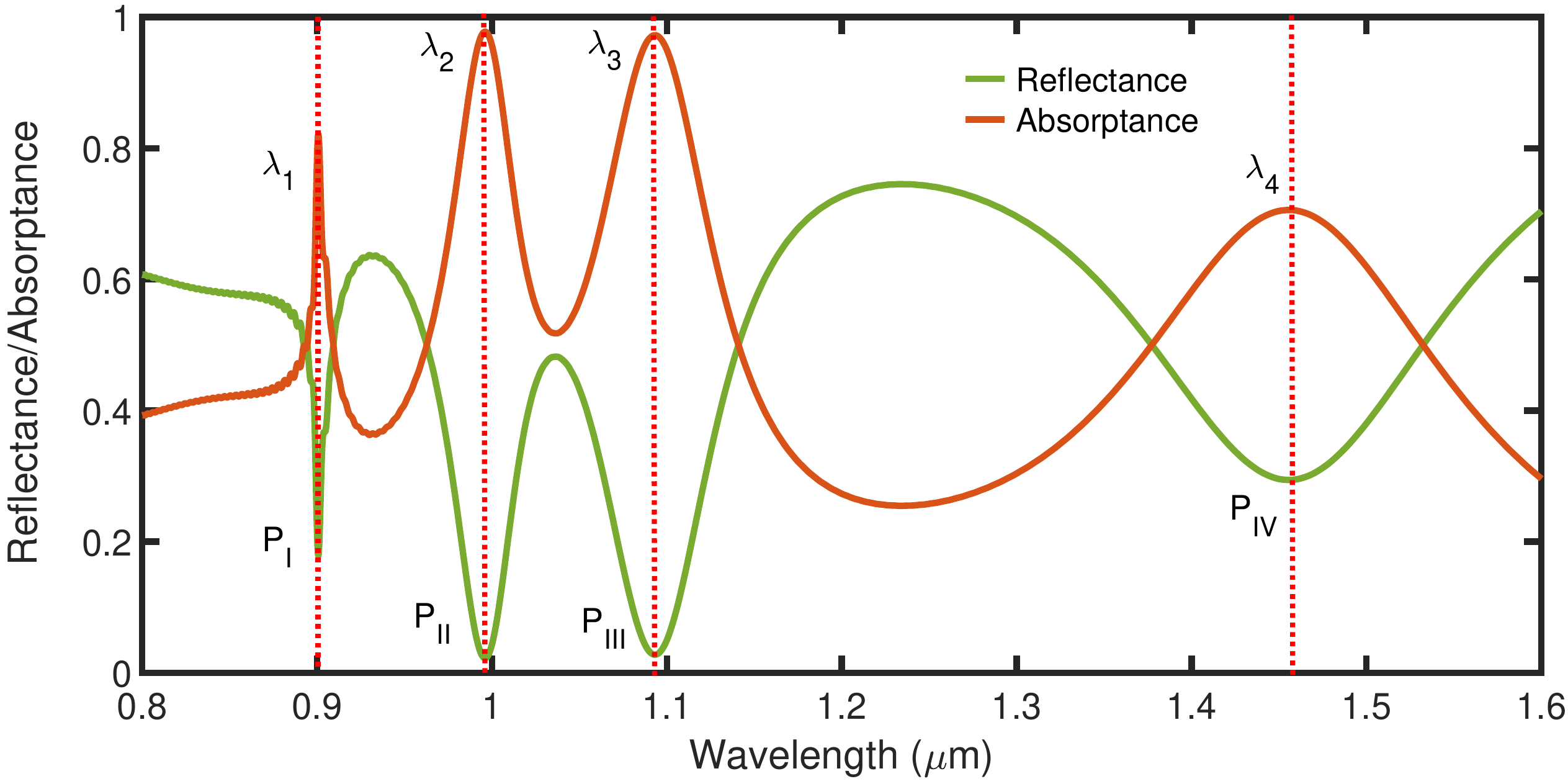}
  \caption{Simulated absorptance and reflectance spectra of the hybrid metasurface for the optimized structural parameters, showing four resonant modes labeled $\mathrm{P}_{\mathrm{I}}$, $\mathrm{P}_{\mathrm{II}}$, $\mathrm{P}_{\mathrm{III}}$, and $\mathrm{P}_{\mathrm{IV}}$, occurring at wavelengths $\lambda_1=900\,\mathrm{nm}$, $\lambda_2=990\,\mathrm{nm}$, $\lambda_3=1075\,\mathrm{nm}$, and $\lambda_4=1465 \,\mathrm{nm}$, respectively.}
  \label{simulated_spectra}
\end{figure}

Fig.~\ref{simulated_spectra} presents the absorptance ($A$) and reflectance ($R$) spectra of the hybrid metasurface under the optimized structural parameters defined previously. Since the bottom $\mathrm{Al}$ layer is optically thick, transmission through the structure is negligible, and the relation $A + R = 1$ holds. Four distinct resonance peaks are observed at wavelengths $\lambda_{1} = 900\,\mathrm{nm}$, $\lambda_{2} = 990\,\mathrm{nm}$, $\lambda_{3}=1075\,\mathrm{nm}$, and $\lambda_{4} = 1465\,\mathrm{nm}$. The resonance at $\lambda_{1}$ exhibits a narrow linewidth with a full-width at half maximum (FWHM) of approximately $4.7\,\mathrm{nm}$ and achieves near-unity absorptance ($A\approx 1$), corresponding to a quality factor of $Q=190$. In addition to this narrowband resonance, three broader absorption bands are observed at $\lambda_{2}$, $\lambda_{3}$, and $\lambda_{4}$, with peak absorptance values of $99.4\%$, $99.3\%$, and $78.5\%$, respectively.

\begin{figure*}[t!]
  \centering
   \subfloat[]{\includegraphics[width=0.23\columnwidth]{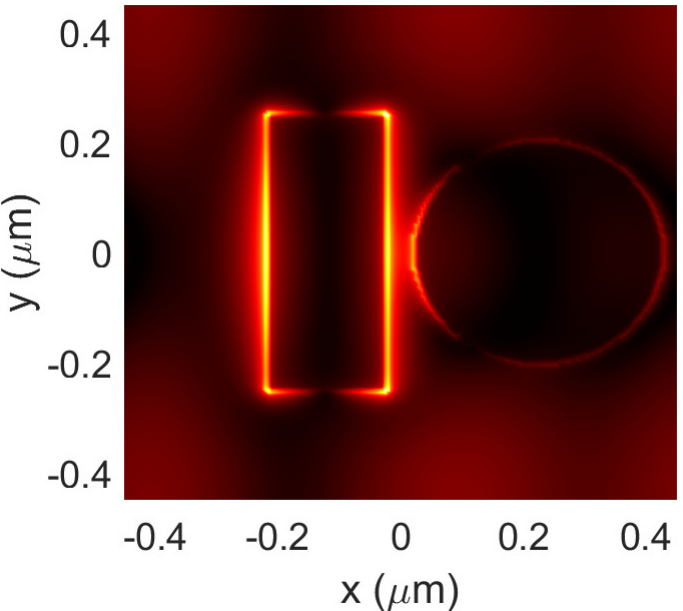}}\hspace{2pt}
   \subfloat[]{\includegraphics[width=0.23\columnwidth]{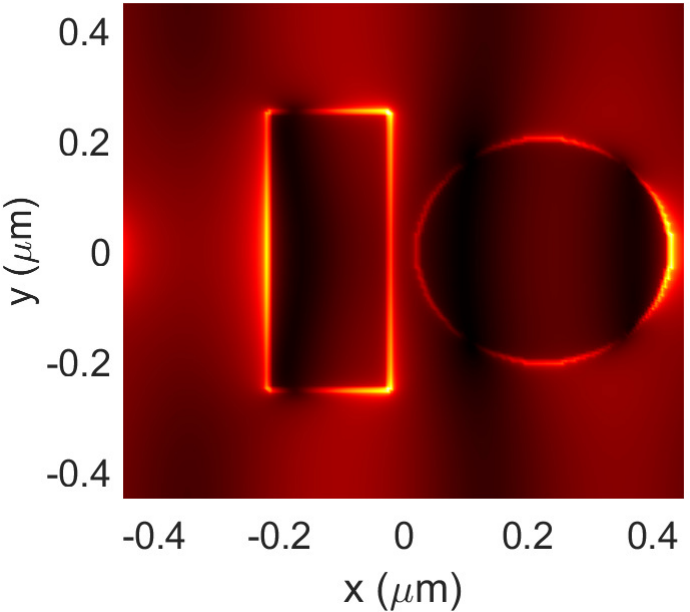}}\hspace{2pt}
   \subfloat[]{\includegraphics[width=0.23\columnwidth]{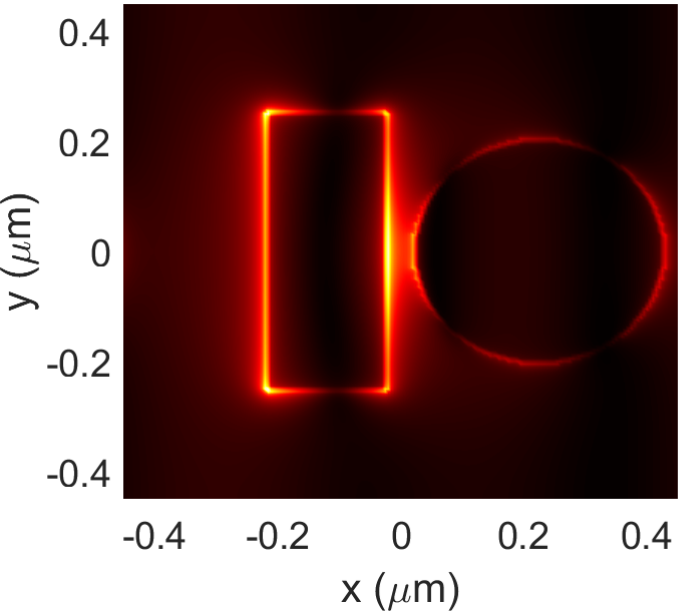}}\hspace{2pt}
   \subfloat[]{\includegraphics[width=0.23\columnwidth]{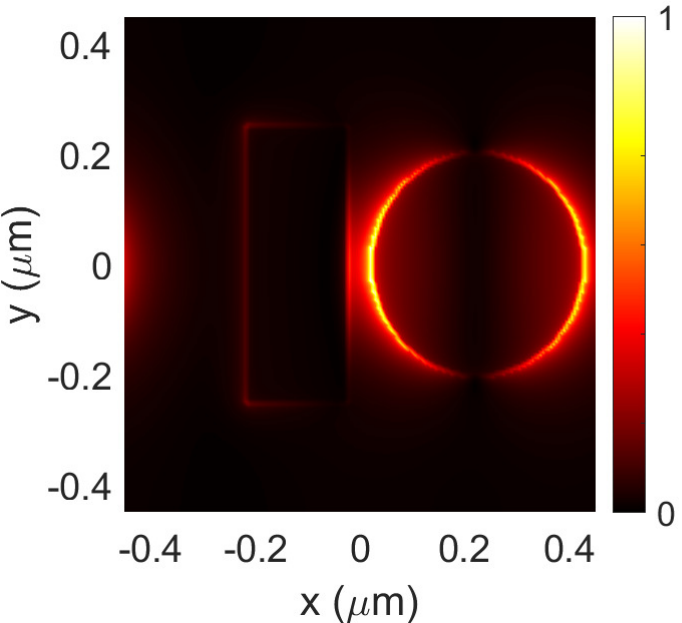}}\\
   \subfloat[]{\includegraphics[width=0.23\columnwidth]{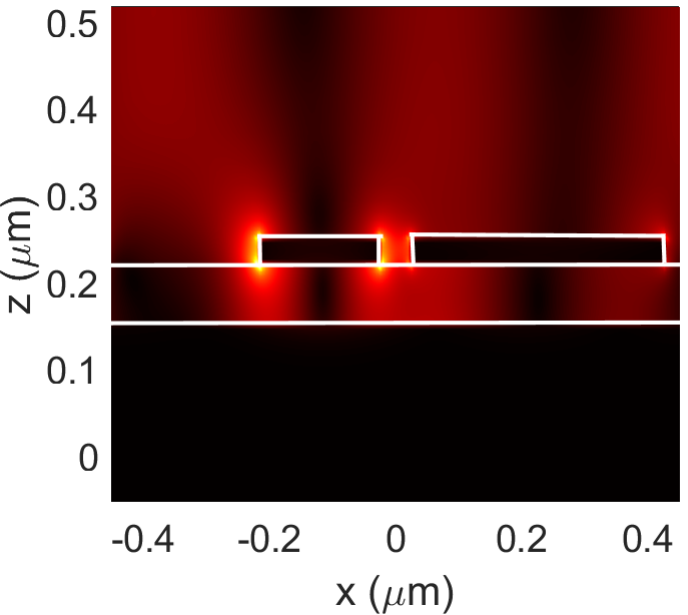}}\hspace{2pt}
   \subfloat[]{\includegraphics[width=0.23\columnwidth]{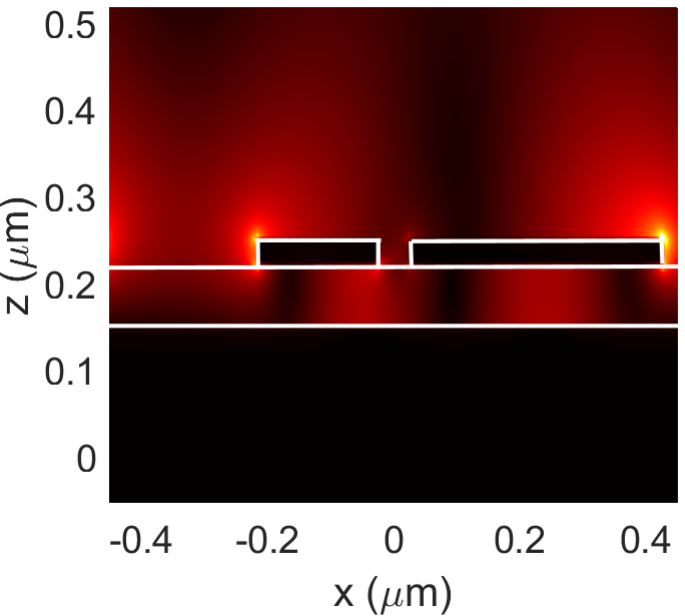}}\hspace{2pt}
   \subfloat[]{\includegraphics[width=0.23\columnwidth]{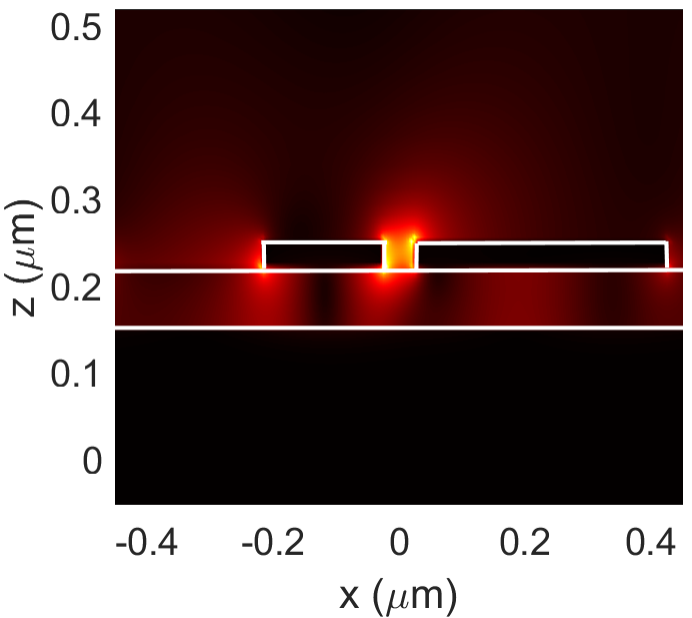}}\hspace{2pt}
   \subfloat[]{\includegraphics[width=0.23\columnwidth]{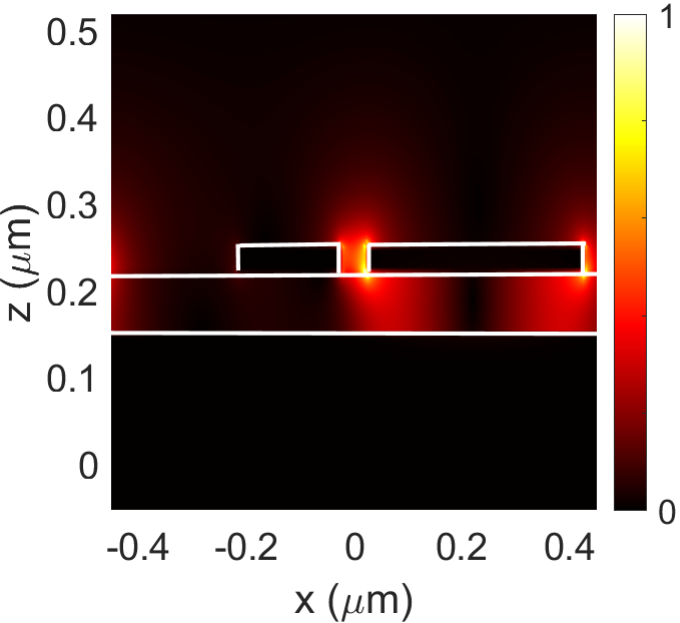}}\\
      \subfloat[]{\includegraphics[width=0.23\columnwidth]{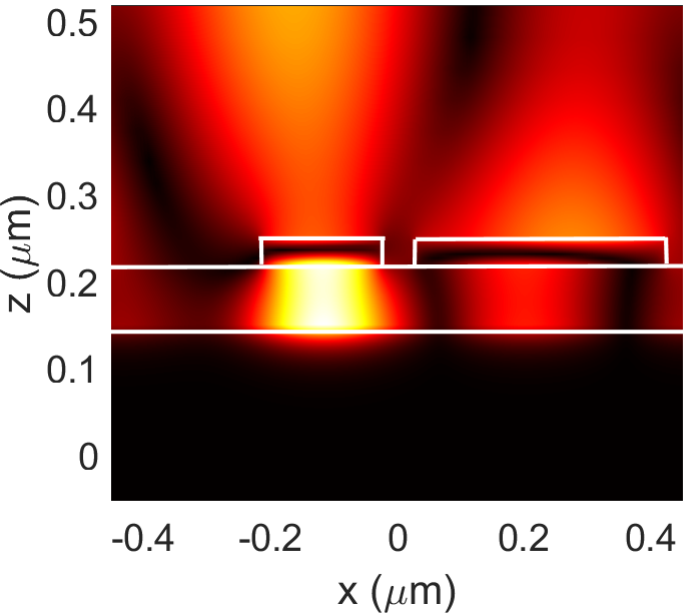}}\hspace{2pt}
      \subfloat[]{\includegraphics[width=0.23\columnwidth]{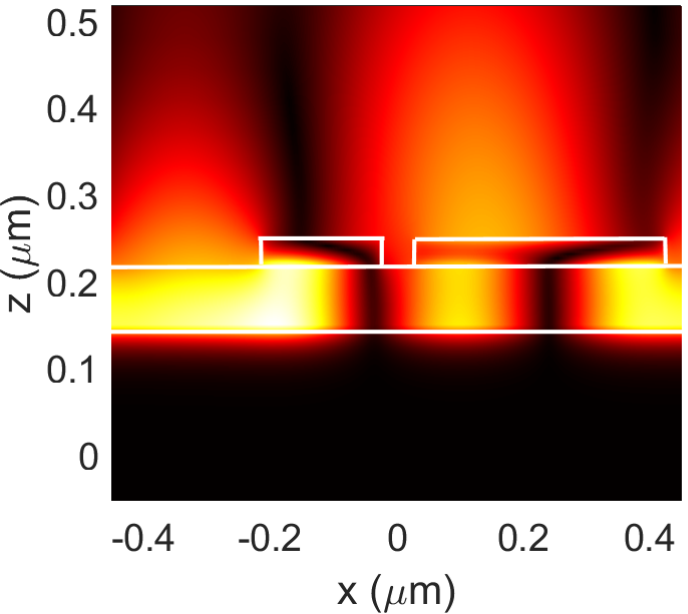}}\hspace{2pt}
      \subfloat[]{\includegraphics[width=0.23\columnwidth]{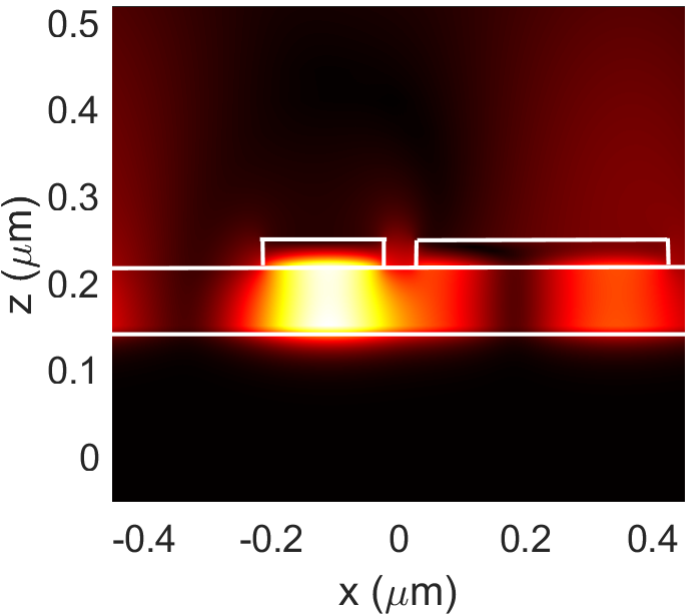}}\hspace{2pt}
      \subfloat[]{\includegraphics[width=0.235\columnwidth]{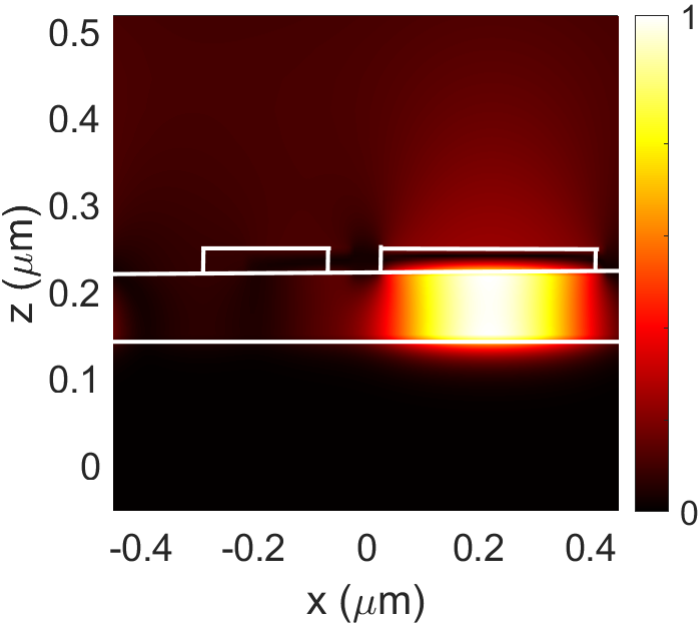}}
  \caption{Electromagnetic field distributions of the hybrid metasurface at the four resonant wavelengths $\lambda_1=900\,\mathrm{nm}$, $\lambda_2=990\,\mathrm{nm}$, $\lambda_3=1075\,\mathrm{nm}$, and $\lambda_4=1465\, \mathrm{nm}$. (a-d) Normalized electric-field distributions in the $xy$-plane at $\lambda_1$, $\lambda_2$, $\lambda_3$, and $\lambda_4$, respectively. (e-h) Corresponding normalized electric-field distributions in the $xz$-plane. (i-l) Corresponding normalized magnetic-field distributions in the $xz$-plane.}
  \label{electric_magnetic_field_plot}
\end{figure*}

To elucidate the origin of the observed multiband resonances, the normalized electric- and magnetic-field distributions of the complete hybrid metasurface are examined at the resonance wavelengths $\lambda_i$ ($i = 1, \dots, 4$), as shown in Fig.~\ref{electric_magnetic_field_plot}. Fig.~\ref{electric_magnetic_field_plot}(a)--\ref{electric_magnetic_field_plot}(d) presents the electric-field distributions in the $xy$-plane, while Fig.~\ref{electric_magnetic_field_plot}(e)--\ref{electric_magnetic_field_plot}(h) and Fig.~\ref{electric_magnetic_field_plot}(i)--\ref{electric_magnetic_field_plot}(l) show the electric- and magnetic-field distributions in the $xz$-plane, respectively.

At $\lambda_{1}$, the electric field in the $xy$-plane is predominantly localized along the bar resonator, as shown in Fig.~\ref{electric_magnetic_field_plot}(a), and the corresponding $xz$-plane profiles [Fig.~\ref{electric_magnetic_field_plot}(e) and~\ref{electric_magnetic_field_plot}(i)] exhibit relatively weak magnetic-field confinement within the dielectric spacer. This field distribution is characteristic of an LSP-dominated mode supported primarily by the bar.

At $\lambda_{2}$, Fig.~\ref{electric_magnetic_field_plot}(b),~\ref{electric_magnetic_field_plot}(f), and~\ref{electric_magnetic_field_plot}(j) shows strong electric- and magnetic-field enhancement both within the dielectric spacer and above the resonators. This behavior indicates the excitation of a hybrid LSP-GSP plasmon mode arising from strong electromagnetic coupling between the bar and disc resonators.

In contrast, at $\lambda_{3}$ and $\lambda_{4}$, the $xz$-plane field distributions [Fig.~\ref{electric_magnetic_field_plot}(g),~\ref{electric_magnetic_field_plot}(h),~\ref{electric_magnetic_field_plot}(k), and~\ref{electric_magnetic_field_plot}(l)] exhibit pronounced confinement of both electric and magnetic fields within the dielectric spacer beneath the resonators. The magnetic-field profiles display a single antinode in the spacer region, confirming the excitation of first-order GSP cavity modes ($p = 1$). The $xy$-plane distributions further indicate that the resonance at $\lambda_{3}$ is primarily associated with the bar, whereas the resonance at $\lambda_{4}$ is predominantly governed by the disc.

\begin{figure*}[t!]
  \centering
  \subfloat[]{\includegraphics[width=0.48\columnwidth]{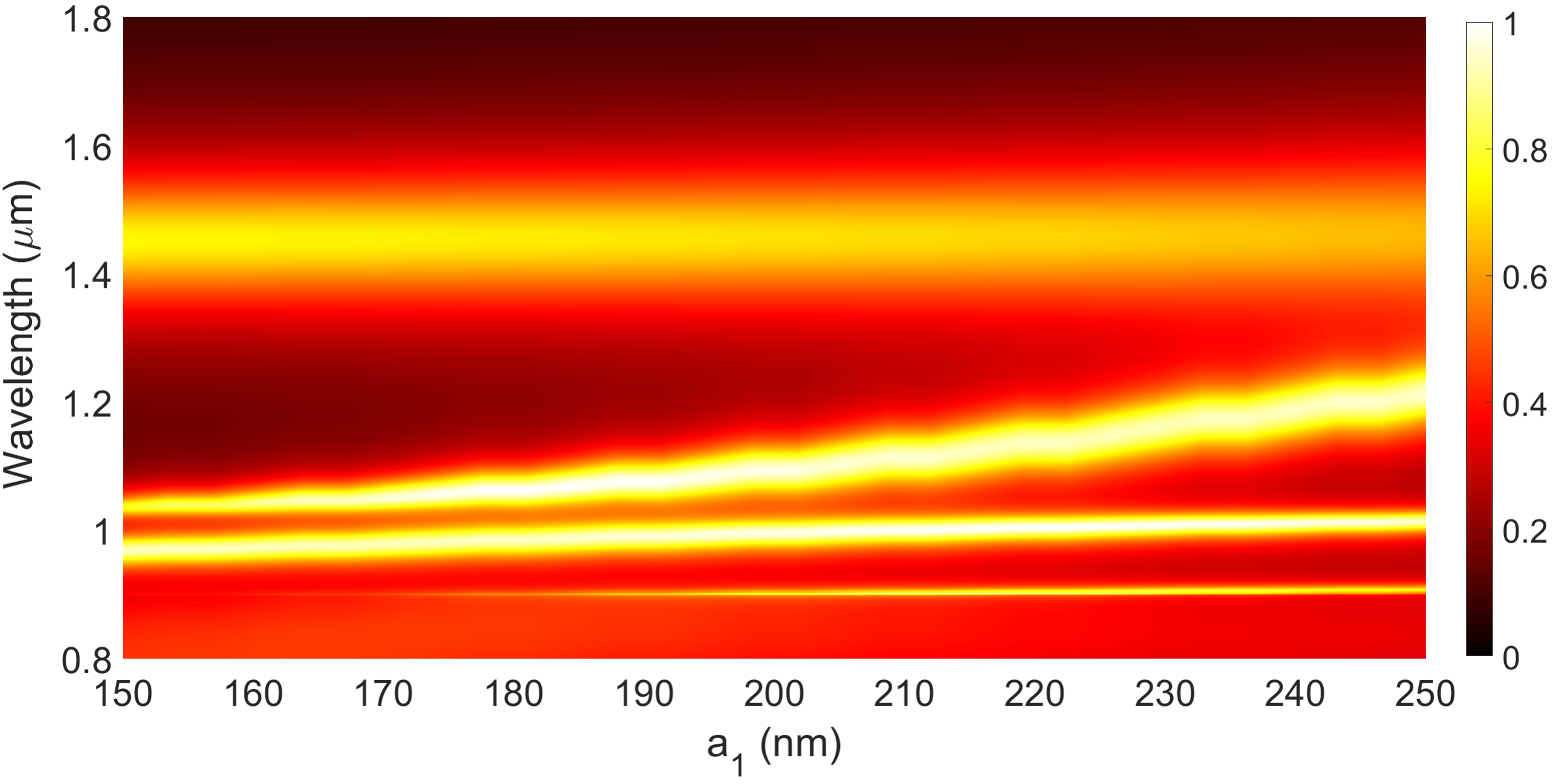}}\hspace{4pt}
  \subfloat[]{\includegraphics[width=0.48\columnwidth]{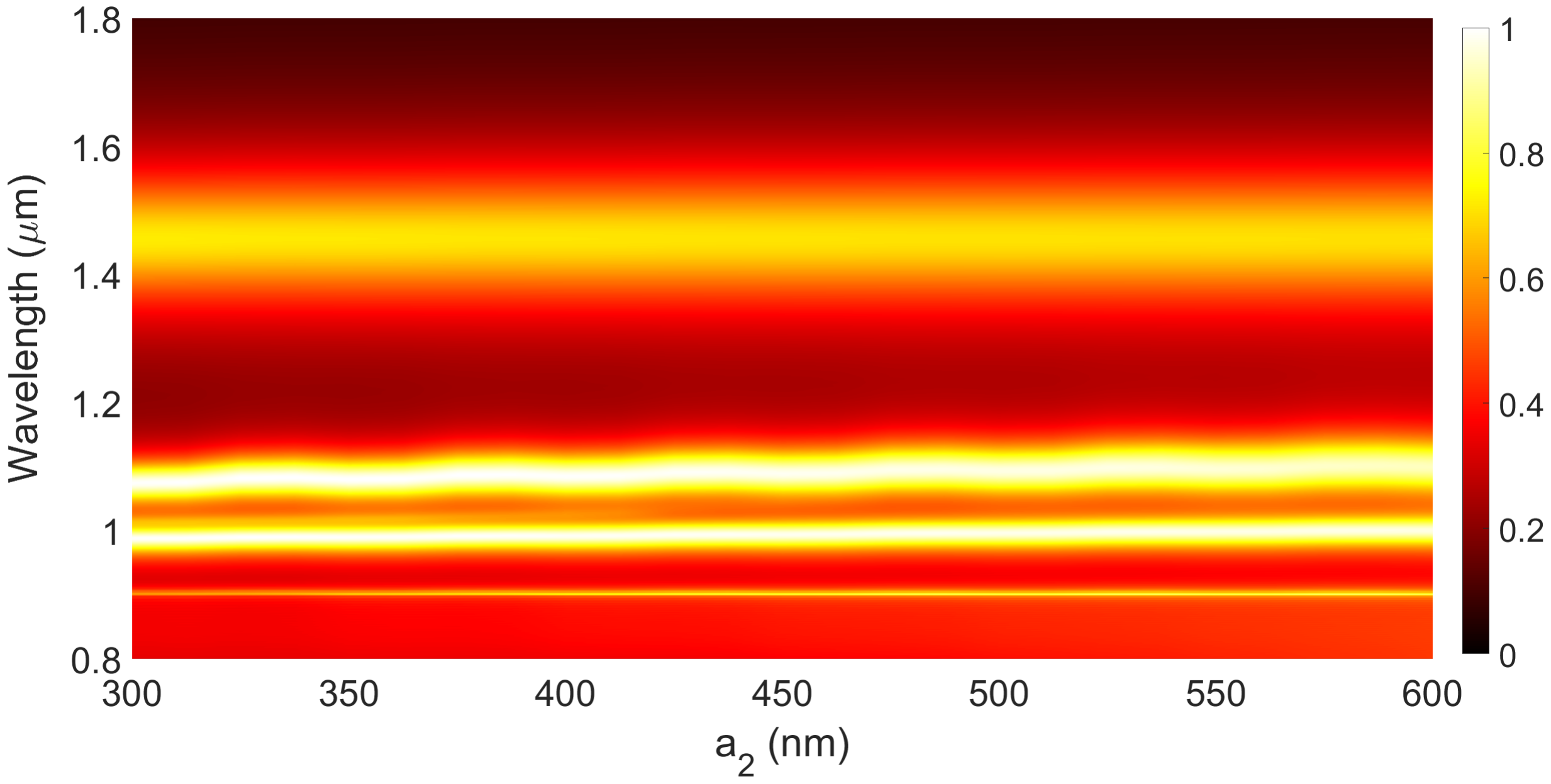}}\\
  \subfloat[]{\includegraphics[width=0.48\columnwidth]{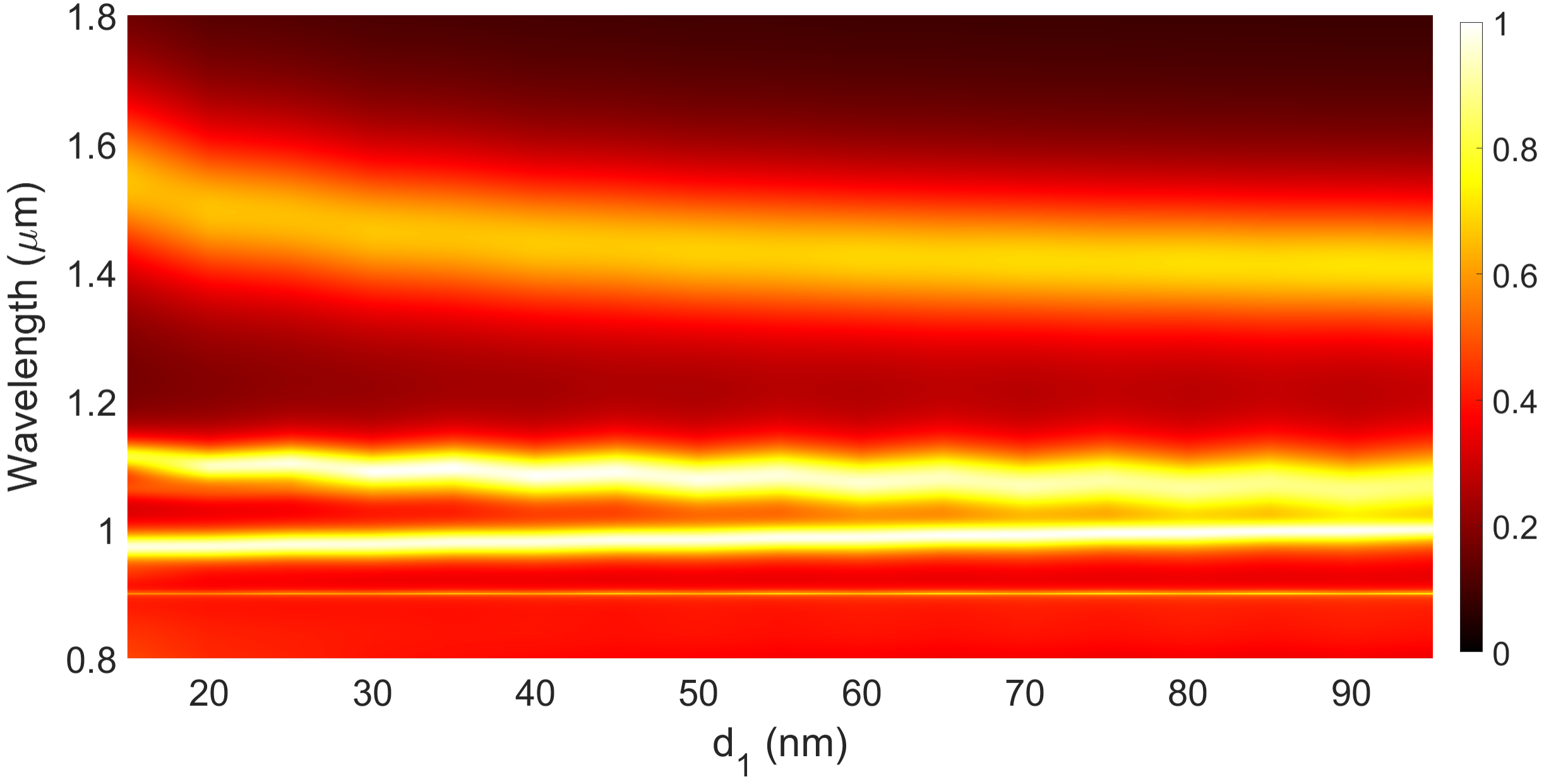}}\hspace{4pt}
  \subfloat[]{\includegraphics[width=0.48\columnwidth]{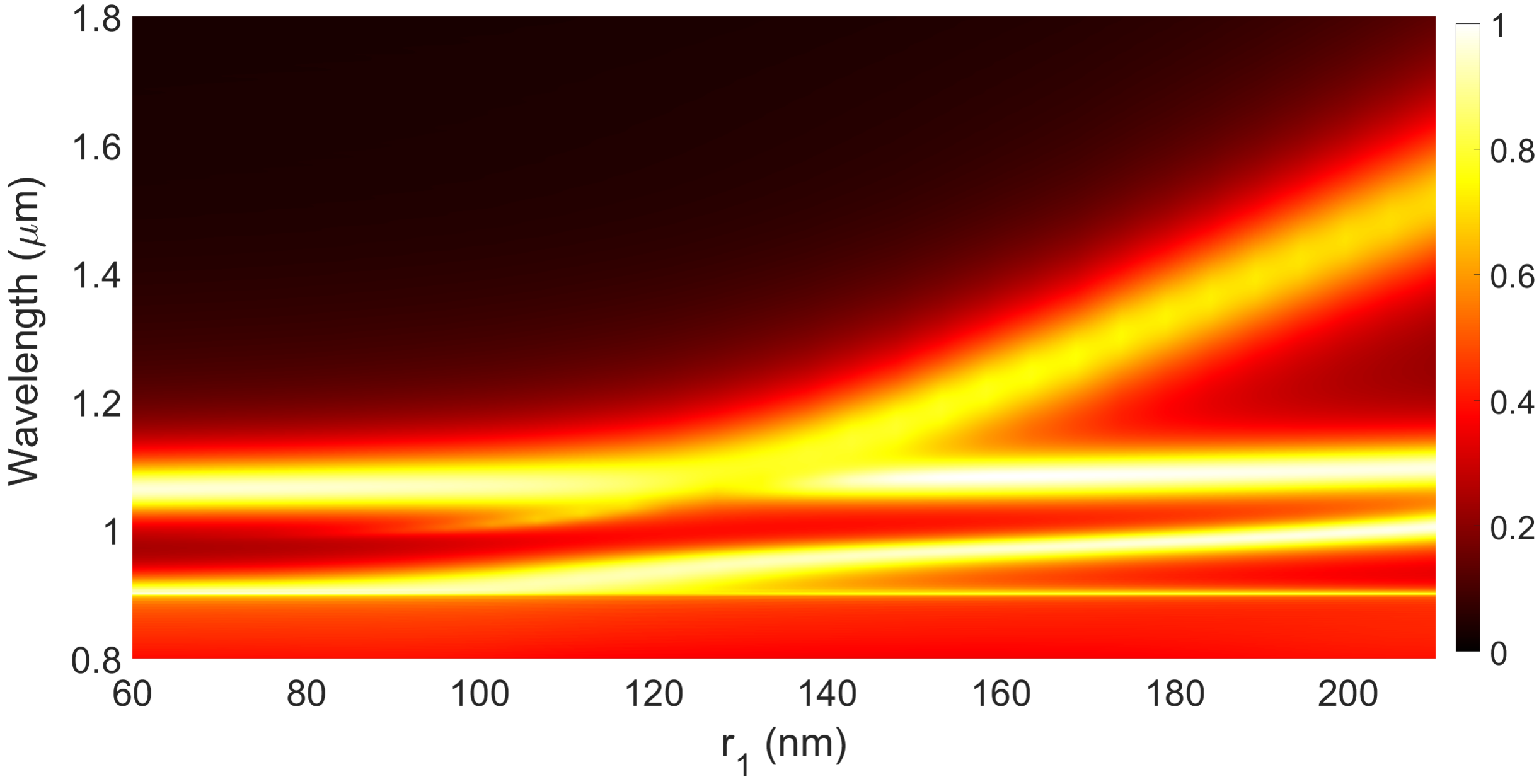}}\\
  \subfloat[]{\includegraphics[width=0.48\columnwidth]{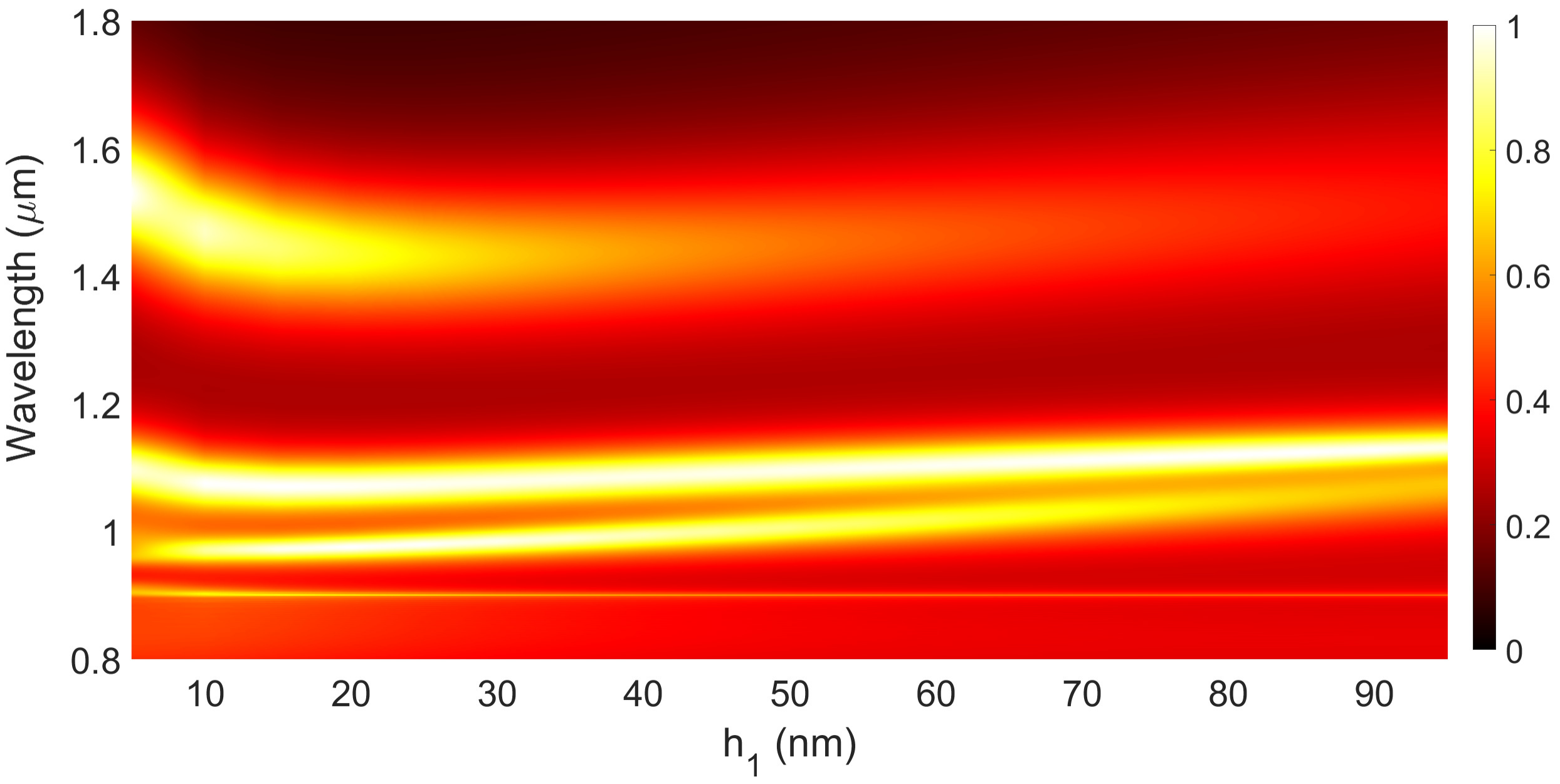}}\hspace{4pt}
  \subfloat[]{\includegraphics[width=0.48\columnwidth]{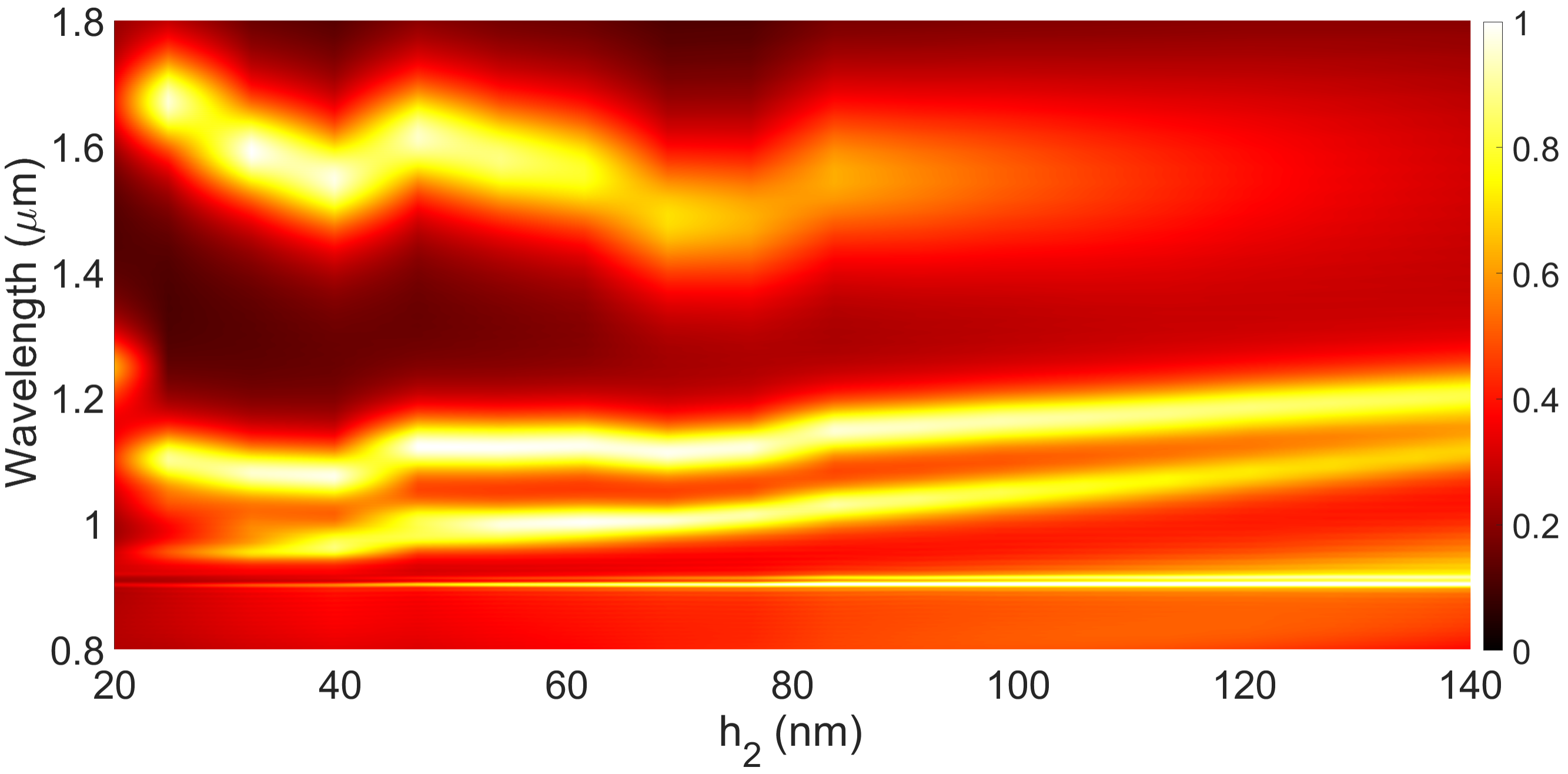}}
  \caption{Simulated absorptance spectra illustrating the influence of geometrical parameters on the hybrid metasurface resonances. (a) Variation of bar width $a_1$. (b) Variation of bar length $a_2$. (c) Variation of bar-disc separation distance $d_1$. (d) Variation of disc radius $r_1$. (e) Variation of top $\mathrm{Al}$ layer thickness $h_1$. (f) Variation of $\mathrm{SiO_2}$ spacer thickness $h_2$.}
   \label{structure_parameters_plot}
\end{figure*}

Next, the influence of the structural parameters on the absorptance spectra is examined, as shown in Fig.~\ref{structure_parameters_plot}. Fig.~\ref{structure_parameters_plot}(a) shows that increasing the bar width $a_1$ results in a redshift of the resonance wavelengths $\lambda_{2}$, $\lambda_{3}$, and $\lambda_{4}$, indicating an increase in the effective lateral cavity length governing the gap-related modes. In contrast, the spectral position of the resonance wavelength $\lambda_1$ remains nearly unchanged with the variations in the bar width $a_{1}$, the bar length $a_{2}$, and the separation distance $d_{1}$, as shown in Fig.~\ref{structure_parameters_plot}(a)--\ref{structure_parameters_plot}(c), respectively. This weak dependence suggests that $\lambda_{1}$ is predominantly associated with a LSP mode supported by the bar.

Variations in the separation distance $d_1$ and the disc radius $r_{1}$, as shown in Fig.~\ref{structure_parameters_plot}(c) and~\ref{structure_parameters_plot}(d), significantly modify the spectral positions of $\lambda_{2}$, $\lambda_{3}$, and $\lambda_{4}$, confirming the role of electromagnetic coupling and gap-related confinement in shaping the hybrid resonances. In particular, the strong sensitivity of $\lambda_{3}$ and $\lambda_{4}$ to geometric variations further supports their identification as first-order GSP cavity modes.

The influence of the top $\mathrm{Al}$ thickness $h_1$ on the absorptance spectra is shown in Fig.~\ref{structure_parameters_plot}(e). Varying $h_1$ produces moderate shifts in the resonance wavelengths, particularly for $\lambda_{2}$, $\lambda_{3}$, and $\lambda_{4}$. This behavior arises from changes in the vertical electromagnetic confinement and the effective coupling between the resonators and the bottom metal layer. In contrast, the resonance at $\lambda_{1}$ exhibits comparatively weak sensitivity to $h_1$, consistent with its LSP character.

The effect of the $\mathrm{SiO_2}$ dielectric spacer thickness $h_2$ is presented in Fig.~\ref{structure_parameters_plot}(f). As $h_2$ increases, the resonance wavelengths $\lambda_{2}$, $\lambda_{3}$, and $\lambda_{4}$ undergo pronounced spectral shifts, reflecting the strong dependence of these resonances on the effective refractive index of the GSP supported by the metal--dielectric--metal configuration. This behavior is consistent with the Fabry--Pérot resonance condition, in which the propagation constant $\beta_{\mathrm{gsp}}$ is highly sensitive to the spacer thickness. In contrast, $\lambda_{1}$ remains only weakly affected by variations in $h_2$, further confirming its LSP nature.

The influence of the bottom $\mathrm{Al}$ thickness $h_3$ on the absorptance spectra is negligible. Since $h_3$ is much larger than the skin depth of aluminum in the near-infrared region, the bottom layer effectively suppresses transmission and does not significantly modify the electromagnetic confinement within the dielectric spacer. Consequently, the resonance wavelengths remain nearly unchanged with variations in $h_3$.

\begin{figure}[t!]
  \centering
   \includegraphics[width=0.85\columnwidth]{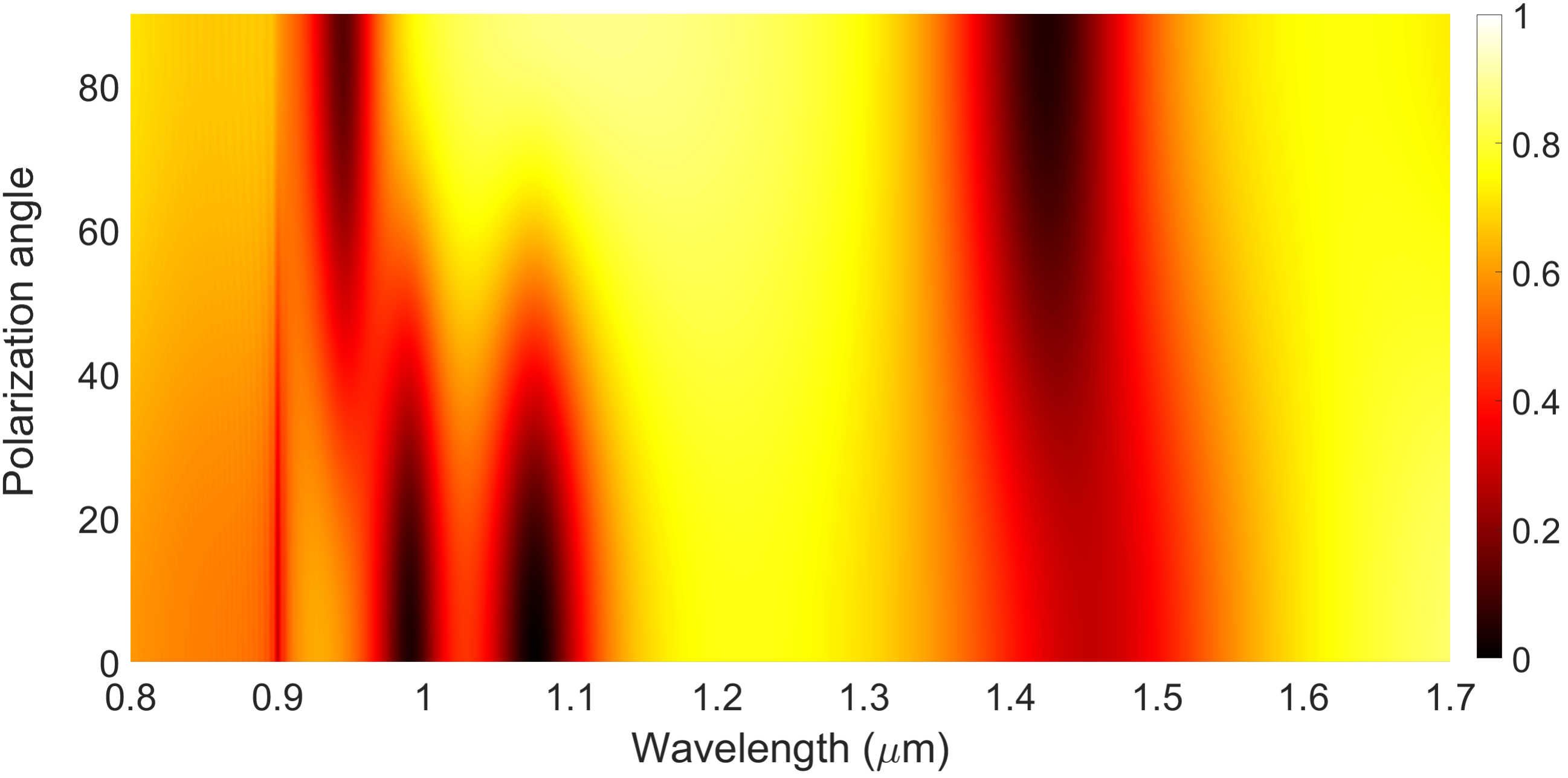}
  \caption{Simulated reflectance spectra of the hybrid metasurface for varying polarization angle $\phi$, defined as the angle between the incident electric field and the $x$-axis in the plane of the metasurface.}
  \label{pol_variation_colorplot}
\end{figure}

To examine the influence of polarization on the reflectance spectra, the polarization angle of the incident electric field is varied from $0^{\circ}$ ($x$-polarized) to $90^{\circ}$ ($y$-polarized). Fig.~\ref{pol_variation_colorplot} shows that the resonance peaks exhibit pronounced sensitivity to the polarization state. As $\phi$ increases from $0^{\circ}$ to $90^{\circ}$, the resonances at $\lambda_{1}$, $\lambda_{2}$, and $\lambda_{3}$ gradually diminish, while a new resonance feature emerges near $930\,\mathrm{nm}$, due to the polarization-dependent excitation of an additional GSP mode supported by the hybrid resonator. Additionally, increasing the polarization angle enhances the absorptance at $\lambda_{4}$, accompanied by a slight blueshift in the resonance wavelength.

\begin{figure}[t!]
  \centering
   \includegraphics[width=0.85\columnwidth]{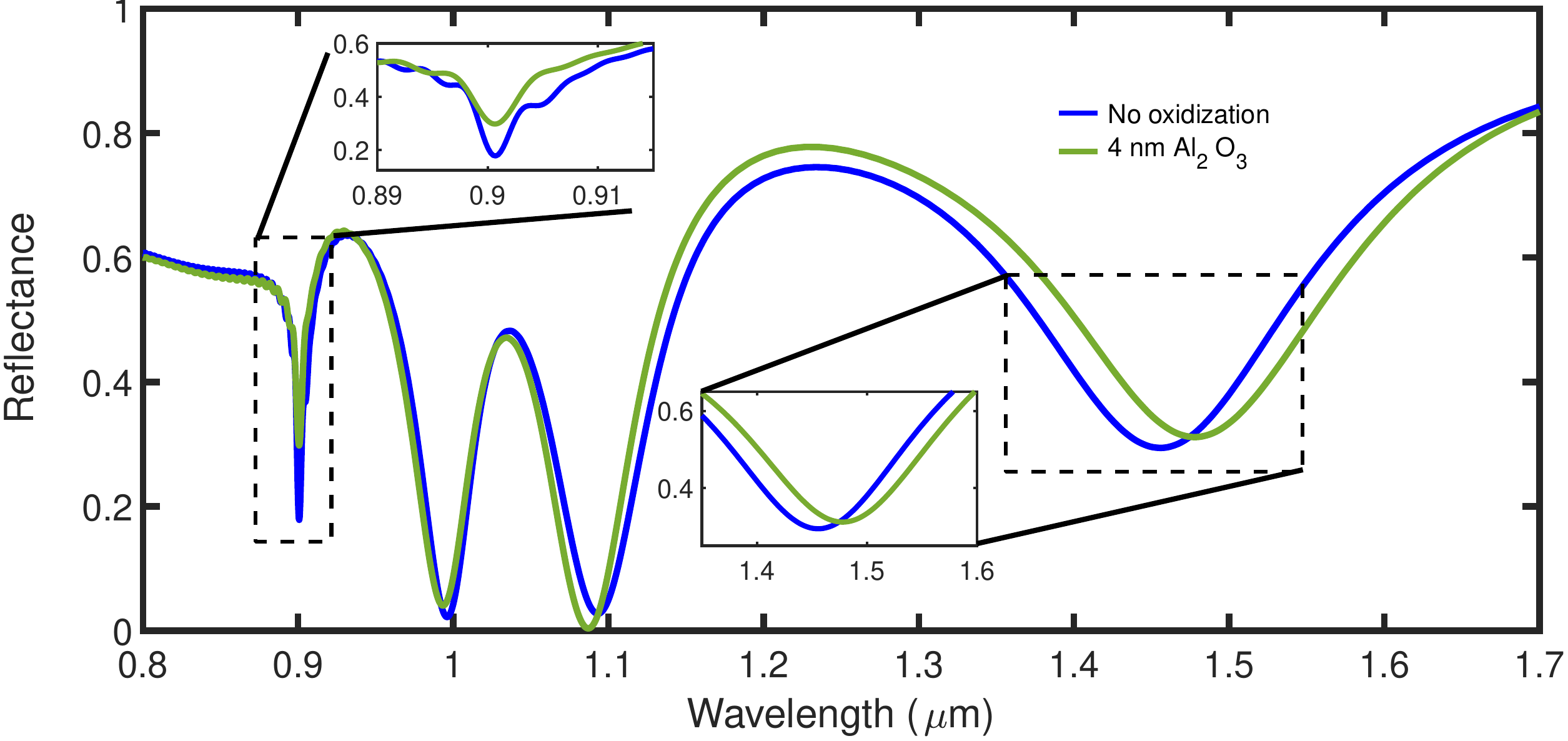}
  \caption{Simulated reflectance spectra of the hybrid metasurface with and without a conformal $\mathrm{Al_2O_3}$ layer. The insets show magnified views of the spectral response near the resonant wavelengths, highlighting the effect of the oxide layer on the linewidth and Q-factor.}
  \label{simulated_spectra_Al2O3}
\end{figure}

In practice, $\mathrm{Al}$ surfaces rapidly form a native oxide layer ($\mathrm{Al_{2}O_{3}}$) with a typical thickness of $2$--$4\,\mathrm{nm}$. To account for this effect, a conformal $4\,\mathrm{nm}$ $\mathrm{Al_{2}O_{3}}$ layer is incorporated in the FDTD simulations, and the resulting reflectance spectra are shown in Fig.~\ref{simulated_spectra_Al2O3}. As evident, the inclusion of the oxide layer results in only a minimal spectral shift of the resonance wavelengths. As further quantified in Table~\ref{tab:table1}, a moderate reduction in both the Q-factor and absorption amplitude is observed across all resonant modes, attributed to the modified local dielectric environment and additional damping introduced by the oxide layer. Nevertheless, the overall spectral response remains largely unchanged, confirming the robustness of the metasurface against the native oxide layer.

\section{Experimental Setup and Results}
\begin{sidewaysfigure}[!ht]
  \centering
   \includegraphics[width=0.85\columnwidth]{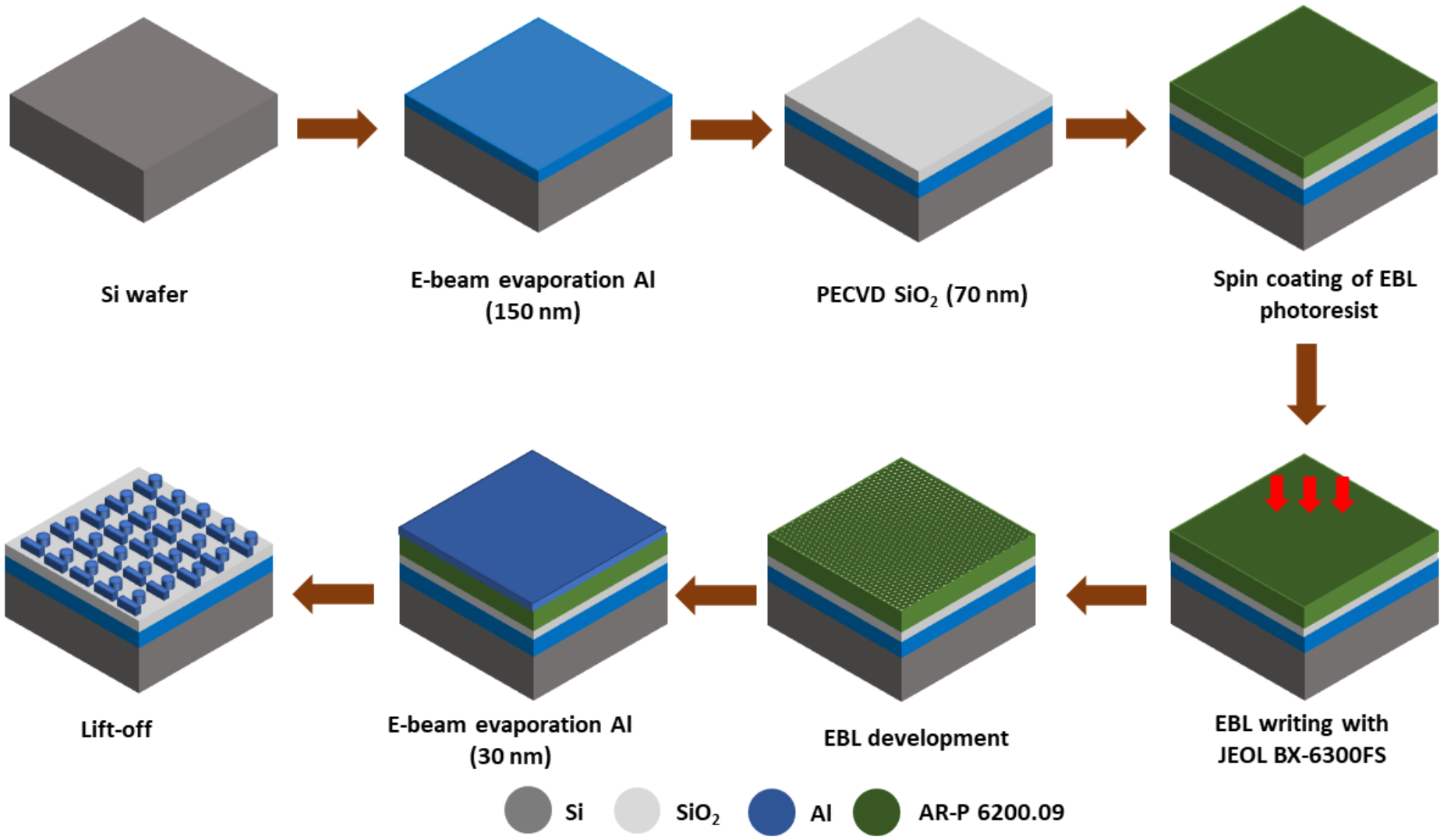}
  \caption{Schematic illustration of the fabrication process flow for the hybrid metasurface.}
   \label{Fab_process}
\end{sidewaysfigure}

\begin{figure}[t!]
  \centering
   \includegraphics[width=0.85\columnwidth]{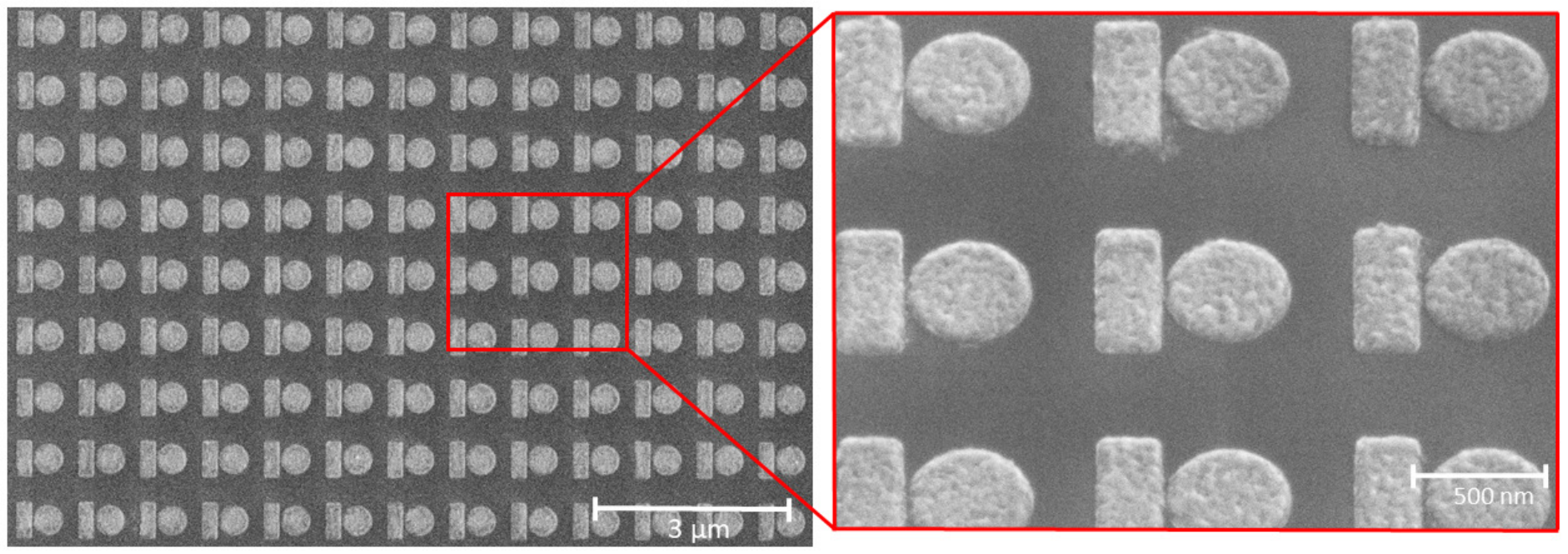}
  \caption{SEM image of the fabricated sample. The inset shows a magnified tilted-view image of the nanostructured hybrid resonators.}
   \label{exp_SEM_fig}
\end{figure}

The metasurface is fabricated on a silicon ($\mathrm{Si}$) wafer using standard electron-beam lithography (EBL) followed by a metal lift-off process. The bottom $\mathrm{Al}$ layer is deposited by sputtering, and the $\mathrm{SiO_2}$ spacer layer is deposited using plasma-enhanced chemical vapor deposition (PECVD). The top $\mathrm{Al}$ layer is patterned using a JEOL JBX-6300FS EBL system and deposited by electron-beam evaporation, followed by lift-off. The detailed fabrication process flow is shown in Fig.~\ref{Fab_process}. The fabricated sample has a footprint of $500 \,\mathrm{\mu m} \times 500\,\mathrm{\mu m}$. Scanning electron microscope (SEM) images of the sample are shown in Fig.~\ref{exp_SEM_fig}, with the inset showing a magnified view of the nanostructures. 

\begin{figure}[t!]
  \centering
   \includegraphics[width=0.85\columnwidth]{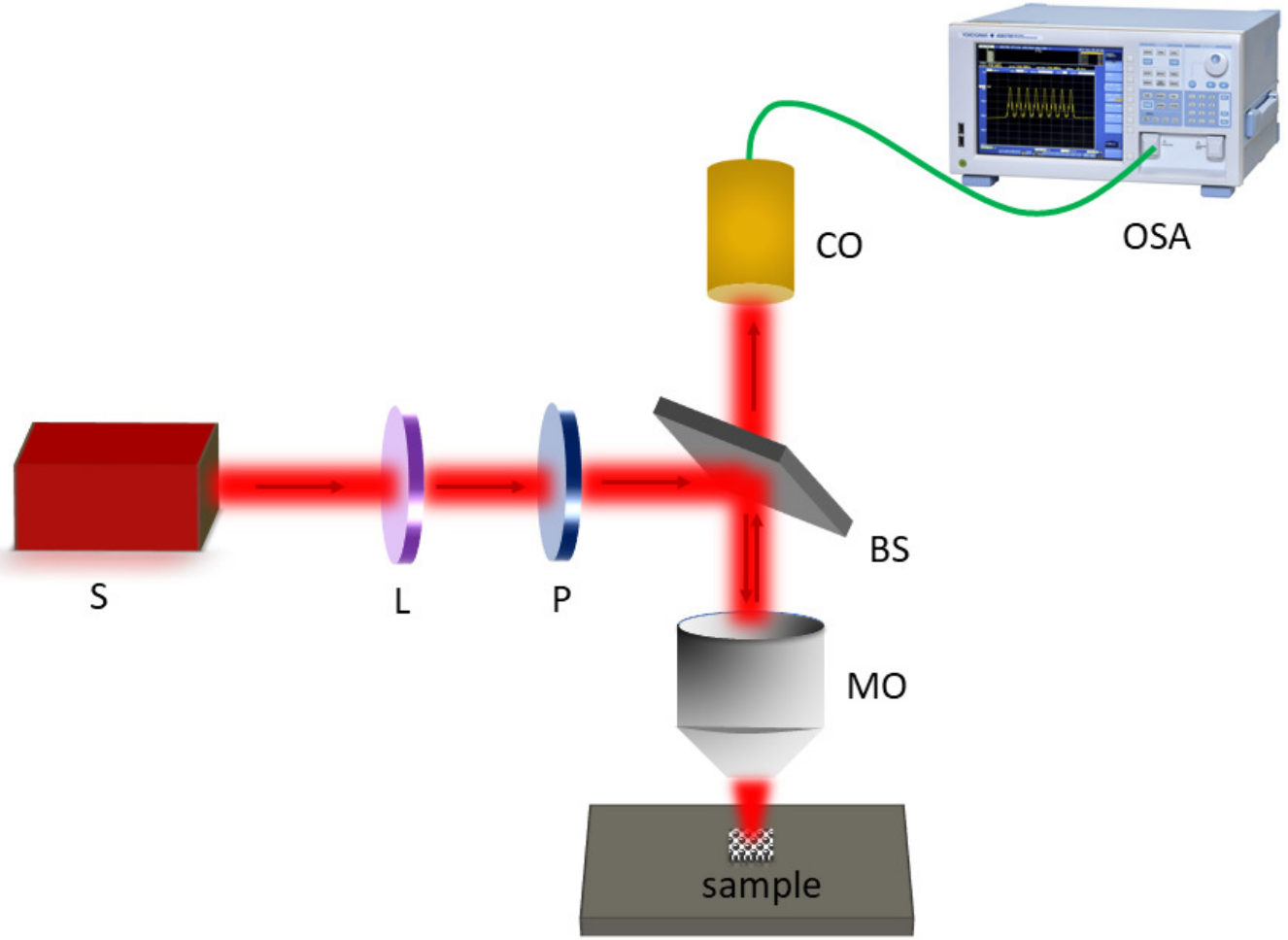}
  \caption{Schematic diagram of the experimental setup used for reflectance measurements. S: broadband source; L: lens; P: broadband polarizer; BS: beam splitter; MO: microscope objective; CO: Collimator; OSA: optical spectrum analyzer.}
   \label{exp_etup_fig}
\end{figure}

For reflectance measurements, a broadband halogen light source (StellarNet Inc. SL1) is used for illumination. A broadband polarizer controls the polarization of the incident light, which is focused onto the sample using a $20\times$ objective lens with a numerical aperture of $0.5$. The reflected light is collected through the same objective and coupled into a multimode optical fiber connected to an optical spectrum analyzer (Yokogawa AQ6370B). A schematic of the experimental setup is shown in Fig.~\ref{exp_etup_fig}. 

\begin{figure}[t!]
  \centering
   \includegraphics[width=0.85\columnwidth]{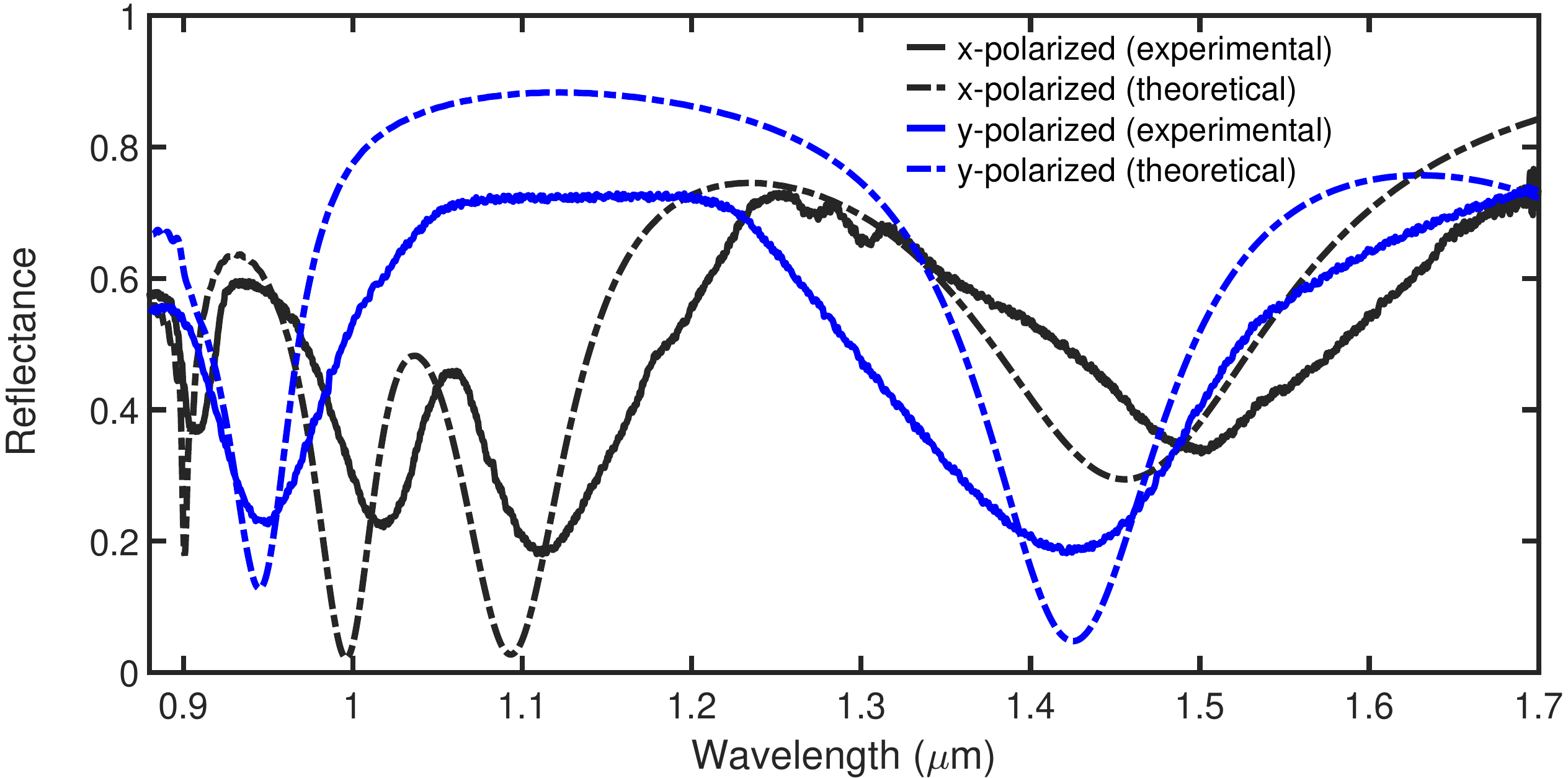}
  \caption{Measured and simulated reflectance spectra of the hybrid metasurface under normal incidence for $x$- and $y$-polarized excitation. The resonance features correspond to the modes $\mathrm{P_I}$--$\mathrm{P_{IV}}$ identified in the simulated spectra.}
   \label{Experimental reflactance}
\end{figure}

The simulated and experimentally measured reflectance spectra are presented in Fig.~\ref{Experimental reflactance} for $x$- and $y$-polarized excitation under normal incidence. For $x$-polarized illumination, the resonances are observed at wavelengths $907\,\mathrm{nm}$, $1015.5\, \mathrm{nm}$, $1110.5\,\mathrm{nm}$, and $1494\,\mathrm{nm}$, corresponding to the modes  $\mathrm{P_{I}}$, $\mathrm{P_{II}}$, $\mathrm{P_{III}}$, and $\mathrm{P_{IV}}$, respectively. For $y$-polarized illumination, the resonances occur at wavelengths $945\,\mathrm{nm}$ and $1430\,\mathrm{nm}$. The experimental results show good agreement with the simulation results. 

A quantitative comparison between the simulated and experimentally measured resonance characteristics is summarized in Table~\ref{tab:table1}. The simulated Q-factors are consistently higher than the experimental values. While the use of periodic boundary conditions is numerically appropriate for modeling an infinite periodic metasurface, it does not capture edge scattering, finite-size effects, and related losses present in experimental samples. Additionally, the simulations assume perfectly smooth material interfaces and an absence of fabrication-induced disorder, whereas the experimental structures inherently suffer from surface roughness, fabrication-induced dimensional variations, grain boundary effects, and the presence of a native $\mathrm{Al_2 O_3}$ layer, all of which introduce additional radiative and non-radiative losses that reduce the measured Q-factor. Similarly, the simulated absorption amplitudes are slightly higher than the experimental values, owing to the neglect of scattering losses and structural imperfections in the numerical model. Despite these differences, the simulated and experimentally measured results are in good quantitative agreement, confirming the validity of the adopted dispersive material model and the robustness of the proposed metasurface design.

\section{Computation of SHG Response}
To evaluate the nonlinear response of the proposed metasurface, a theoretical analysis of SHG is performed. 
The enhanced SHG at the resonant wavelengths arises from strong electromagnetic field localization supported by the LSP and GSP modes. 
Under the electric-dipole approximation, SHG is forbidden in centrosymmetric bulk media~\cite{wang2025enhanced,maekawa2020wavelength}. 
In plasmonic nanostructures, however, the dominant SH polarization originates from surface regions where inversion symmetry is inherently broken~\cite{huttunen2019efficient}. 
The second-order nonlinear surface polarization is expressed as~\cite{kruk2015enhanced}
\begin{equation}
\mathbf{P}^{(2)}_{\mathrm{s}}(\mathbf{r},2\omega)
=\boldsymbol{\chi}^{(2)}_{\mathrm{s}} :
\mathbf{E}(\mathbf{r},\omega)
\mathbf{E}(\mathbf{r},\omega)
\label{eqn:SHG_general}
\end{equation}
where $\mathbf{E}(\mathbf{r},\omega)$ denotes the electric field at frequency $\omega$ evaluated at a point $\mathbf{r}$ on the metal surface.

The surface nonlinear susceptibility tensor $\boldsymbol{\chi}^{(2)}_{\mathrm{s}}$ 
consists of three independent components: 
$\chi^{(2)}_{\mathrm{s},\perp\perp\perp}$, 
$\chi^{(2)}_{\mathrm{s},\perp\parallel\parallel}$, and 
$\chi^{(2)}_{\mathrm{s},\parallel\parallel\perp}
=
\chi^{(2)}_{\mathrm{s},\parallel\perp\parallel}$, 
where $\perp$ and $\parallel$ denote the directions normal and tangential to the surface, respectively~\cite{ethis2016wave}. 
It is well established that the dominant contribution to SHG arises from the normal component of the nonlinear polarization~\cite{yang2017enhancement}. 
Accordingly, only $\chi^{(2)}_{\mathrm{s},\perp\perp\perp}$ is considered in the present simulations. Projecting~\eqref{eqn:SHG_general} onto the surface normal direction yields
\begin{equation}
P^{(2)}_{\mathrm{s},\perp}(\mathbf{r},2\omega)
=
\chi^{(2)}_{\mathrm{s},\perp\perp\perp}
E_{\perp}^{2}(\mathbf{r},\omega)
\label{eqn:SHG_reduced}
\end{equation}
where $E_{\perp}(\mathbf{r},\omega)$ denotes the normal component of the electric field at frequency $\omega$ evaluated at the surface point $\mathbf{r}$.

The SHG response is computed using a nonlinear scattering framework based on the Lorentz reciprocity principle~\cite{rahimi2018nonlinear}. Within this framework, the far-field second-harmonic electric field is obtained, up to a proportionality constant, from the overlap integral over the metal surface $S$~\cite{sugita2023surface,sun2024230},
\begin{equation}
\mathbf{E}_{\mathrm{SH}}(2\omega)
\propto
\int_{S}
\chi^{(2)}_{\mathrm{s},\perp\perp\perp}
E_{\perp}^{2}(\mathbf{r},\omega)
E^{\mathrm{adj}}_{\perp}(\mathbf{r},2\omega)
\, \mathrm{d}S
\label{eqn:overlap}
\end{equation}
where $E^{\mathrm{adj}}_{\perp}(\mathbf{r},2\omega)$ denotes the normal component of the adjoint electric field at frequency $2\omega$, generated by an excitation launched from the detector and evaluated at the surface point $\mathbf{r} \in S$.

\begin{figure}[t!]
\centering
\subfloat[]{\includegraphics[width=0.465\columnwidth]{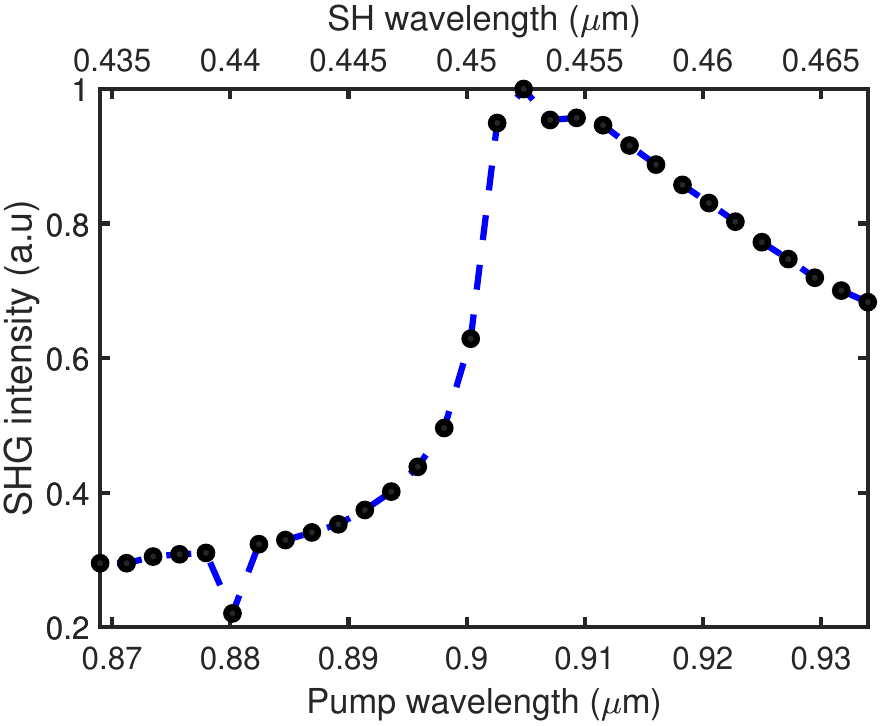}}\hspace{4pt}
\subfloat[]{\includegraphics[width=0.48\columnwidth]{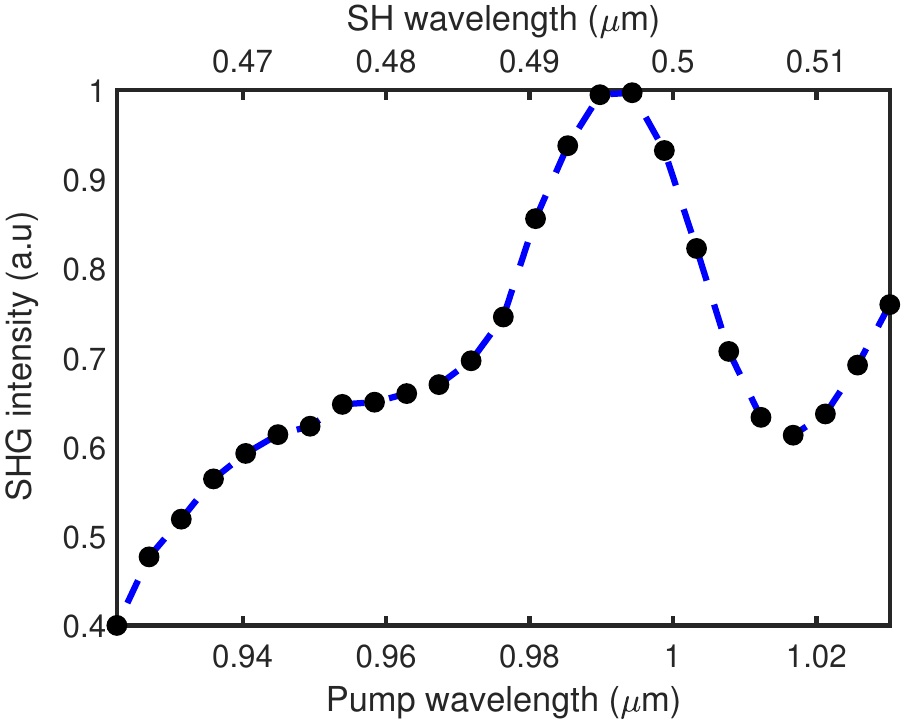}}\\
\subfloat[]{\includegraphics[width=0.48\columnwidth]{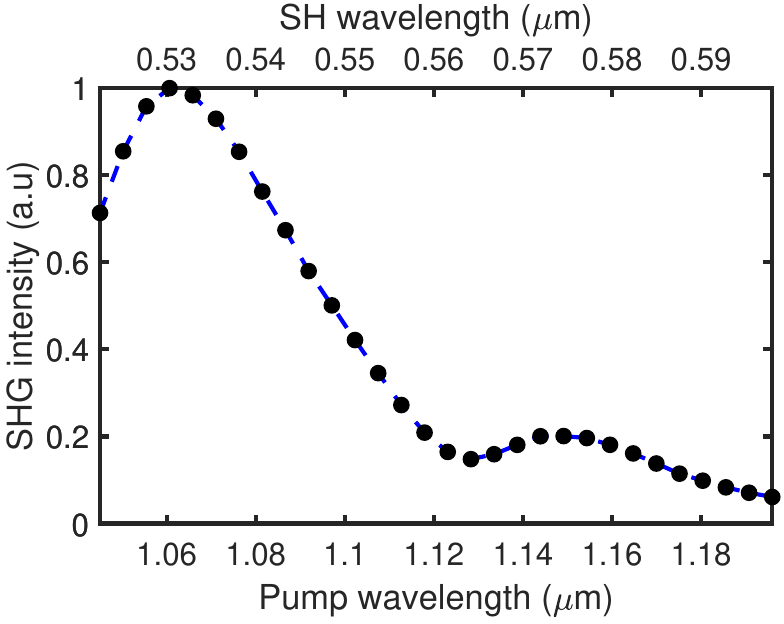}}\hspace{4pt}
\subfloat[]{\includegraphics[width=0.48\columnwidth]{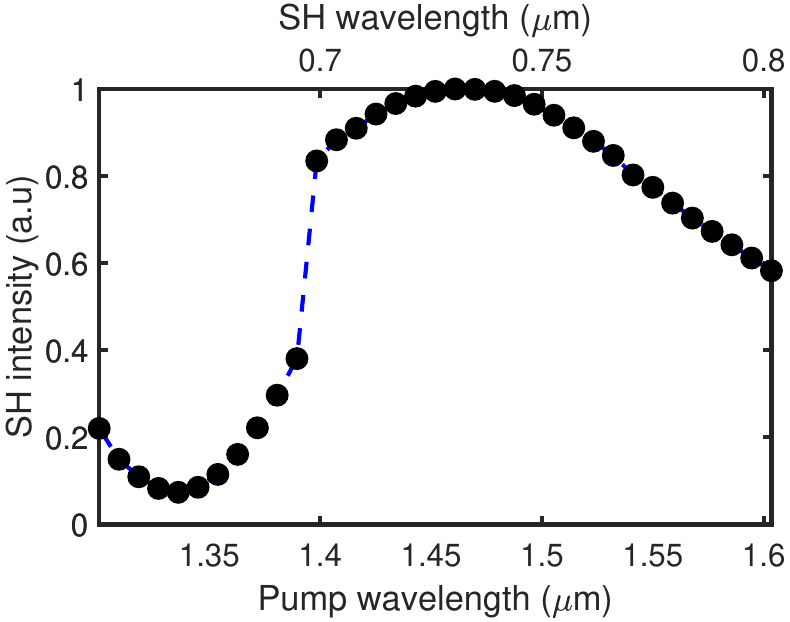}}
\caption{Simulated SHG intensity as a function of the fundamental wavelength for a peak excitation intensity of $2.85~\mathrm{MW/cm^{2}}$. (a–d) SHG responses corresponding to the resonant modes $\mathrm{P_I}$, $\mathrm{P_{II}}$, $\mathrm{P_{III}}$, and $\mathrm{P_{IV}}$, respectively.}
\label{SHG_pump wavelength variation}
\end{figure}

The SHG response of the hybrid metasurface is simulated under normal incidence of the fundamental wave, consisting of $200$ $\mathrm{fs}$ pulses at an $80\,\mathrm{MHz}$ repetition rate. The dependence of the SHG intensity on the fundamental wavelength is analyzed under $x$-polarized femtosecond excitation for a peak pump intensity of $2.85\,\mathrm{MW/cm^{2}}$. The variations in SHG intensity corresponding to the resonant modes $\mathrm{P_I}$–$\mathrm{P_{IV}}$ are presented in Fig.~\ref{SHG_pump wavelength variation}(a)--\ref{SHG_pump wavelength variation}(d). The SHG signal exhibits clear enhancement in the vicinity of the resonance wavelengths, consistent with the strong field localization associated with these modes, as discussed in Section~\ref{metasurface_simulation}. 

\begin{figure}[t!]
  \centering
   \includegraphics[width=0.85\columnwidth]{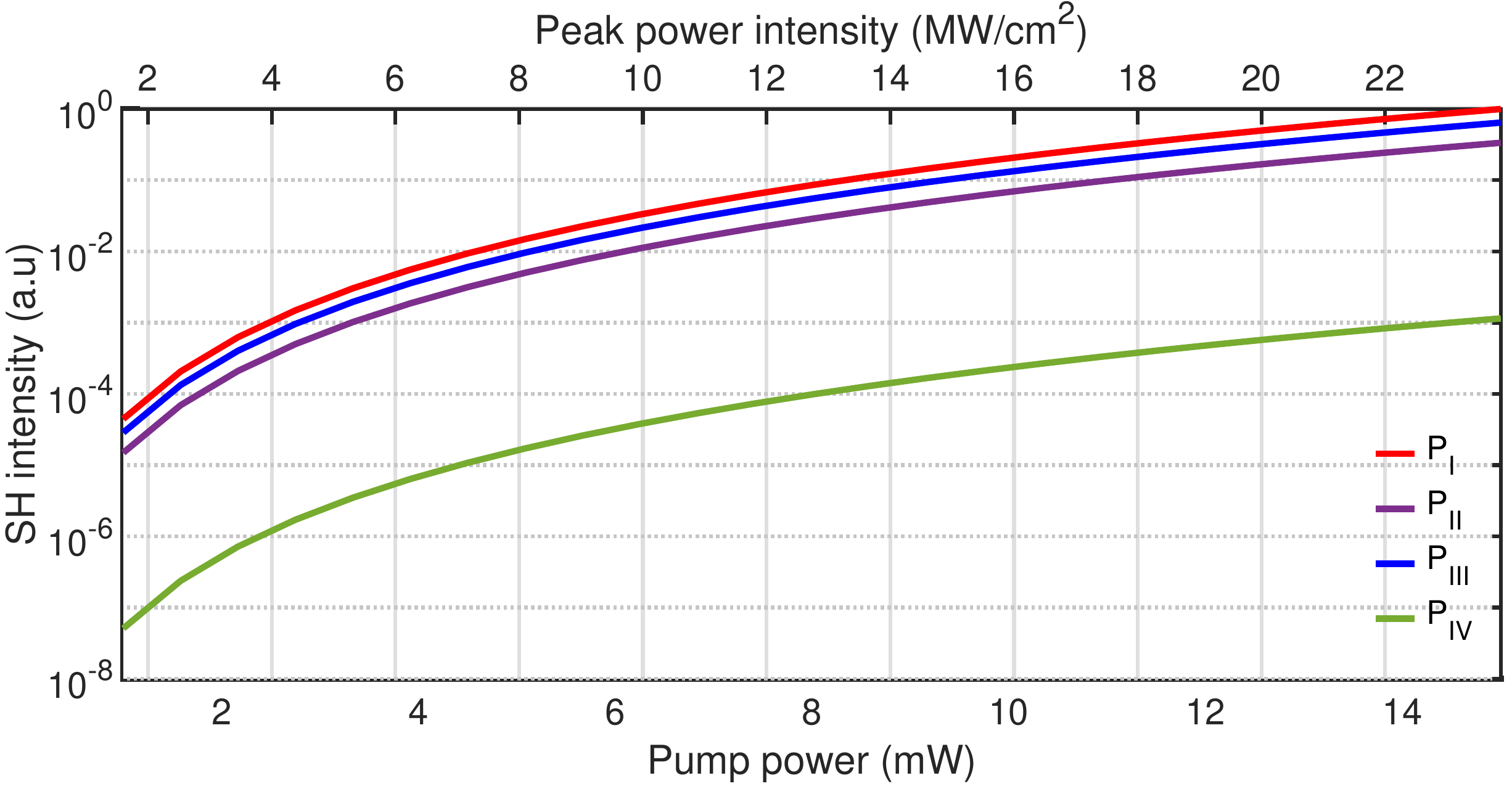}
  \caption{Simulated SHG intensity as a function of the input pump power at the resonant wavelengths.}
   \label{power_variation_combined}
\end{figure}

The SHG conversion efficiency is defined as $P_{2 \omega} / P_\omega$, where $P_{2 \omega}$ denotes the total radiated SH power obtained from the nonlinear scattering formulation and $P_\omega$ is the incident peak pump power under the specified pulsed excitation conditions. The calculated SHG conversion efficiencies are on the order of $10^{-12}$ for modes $\mathrm{P_I}$ and $\mathrm{P_{III}}$, $10^{-13}$ for $\mathrm{P_{II}}$, and approximately $10^{-15}$ for $\mathrm{P_{IV}}$ for the peak pump intensity of $2.85\,\mathrm{MW/cm^{2}}$. The dependence of the SHG intensity on the excitation power at resonant wavelengths is shown in Fig.~\ref{power_variation_combined}. For average powers below $3\,\mathrm{mW}$, the SHG intensity follows a quadratic dependence on the input power, yielding a slope of approximately $2$ on a log–log scale, consistent with a second-order nonlinear process. At higher excitation powers, deviations from the quadratic behavior are observed, indicating the onset of saturation. This behavior is attributed to saturation of the nonlinear polarization at high excitation intensities~\cite{raygoza2019polarization}. 

Among the considered modes, the maximum SHG intensity is obtained at the LSP-dominated $\mathrm{P_{I}}$ resonance, whereas the minimum response occurs at the $\mathrm{P_{IV}}$ mode. The relatively weaker SHG emission at $\mathrm{P_{IV}}$ is associated with the GSP resonance of the disc resonator, whose more symmetric geometry leads to reduced far-field SH radiation efficiency.

To benchmark the proposed design against recently reported GSP-based multiband metasurfaces, a comprehensive comparison is provided in Table~\ref{tab:table2}. The comparison demonstrates that, in contrast to many prior works predominantly limited to dual-band operation or purely numerical results, the proposed metasurface achieves multiresonant behavior with experimental validation and enhanced nonlinear functionality through SHG.

\section{Conclusion}
A multiband GSP-based metasurface incorporating a bar--disc hybrid resonator has been demonstrated to support four distinct resonances across the near-infrared and telecommunication bands. These resonances arise from the controlled interplay between GSP and LSP modes, enabling multiband operation within a single compact meta-atom. The resonance characteristics are tunable through structural parameters and polarization, and experimental measurements show good agreement with numerical predictions, validating the proposed design strategy. Furthermore, numerical analysis of the nonlinear response reveals enhanced SHG at the resonant wavelengths due to strong electromagnetic field localization within the metal--dielectric--metal cavity. The hybrid metasurface provides a versatile platform for multiband and multifunctional nanophotonic devices requiring extended frequency coverage and enhanced nonlinear light–matter interactions.

\section*{Acknowledgments}
The authors acknowledge the KAUST Nanofabrication Core Lab for providing the fabrication facilities.


\bibliographystyle{IEEEtran}
\bibliography{sample_final}

\newpage\clearpage
\section*{Figures and Tables}

\newpage\clearpage

\newpage\clearpage

\newpage\clearpage

\newpage\clearpage

\newpage\clearpage

\newpage\clearpage

\newpage\clearpage

\newpage\clearpage

\newpage\clearpage

\newpage\clearpage

\newpage\clearpage

\newpage\clearpage
\begin{sidewaystable}[!t]
\renewcommand{\arraystretch}{1.25}
\centering
\small
\caption{Quantitative Comparison of Simulated and Experimentally Measured Q-factors and Absorption Amplitudes for All Resonant Modes}\label{tab:table1}
\begin{tabular}{c|c|c|c|c|c|c}
Resonance Mode & $Q$ (Sim.) & \makecell{$Q$ (Sim.) \\with $\mathrm{Al_{2}O_{3}}$} & $Q$ (Exp.) & Abs. (Sim.) ($\%$) & \makecell{Abs. (Sim.) \\ with $\mathrm{Al_{2}O_{3}}$ ($\%$)} & Abs. (Exp.) ($\%$) \\

\hline
$\mathrm{P_{I}}$ & $190$ & $140$ & $75$  & $82$ & $70$ & $63$\\
\hline

$\mathrm{P_{II}}$ & $32$ & $30$ & $26$ & $98$ & $96$ & $80$\\
\hline

$\mathrm{P_{III}}$ & $25$ & $22$ & $17$ & $97$ & $99$ & $82$\\
\hline

$\mathrm{P_{IV}}$ & $10$ & $9$ & $8$ & $71$ & $69$ & $66$\\
\end{tabular}
\end{sidewaystable}
\newpage\clearpage

\begin{sidewaystable}[!ht]
\renewcommand{\arraystretch}{2.5}
\centering
\scriptsize
\caption{Performance Comparison of GSP-based Plasmonic Multiband Metasurfaces}\label{tab:table2}
\begin{tabular}{c|c|c|c|c|c|c}
Ref.  & Structure Type  & Mechanism &  $\#$ Bands & \makecell{Spectral Range} ($\mathrm{nm}$) & Key Contribution & Key Limitation\\
\hline
~\cite{deshpande2019dual} & \makecell {MDM GSP metasurface \\ ($\mathrm{Au}$/$\mathrm{SiO_{2}}$/$\mathrm{Au}$)} & \makecell{Multi-order GSP\\ resonance} & 2 & $633$, $1450$ & \makecell{Phase-gradient \\GSP metasurface} & \makecell{Limited to dual-band,\\ low Q}\\
\hline

~\cite{thrane2022mems} & \makecell {Tunable GSP metasurface\\ $\mathrm{Au}$/air/$\mathrm{Au}$} & \makecell{GSP + Fabry–Perot \\ hybrid} & 1 & $800$ & \makecell{Dynamic blazed \\grating} & \makecell{Limited multiband \\capability}\\
\hline
~\cite{tang2018polarization} & \makecell {MDM GSP metasurface \\ ($\mathrm{Ag}$/$\mathrm{SiO_{2}}$/$\mathrm{Ag}$)} & \makecell{Multi-order GSP\\ resonance} & 2 & $850$, $1550$ & \makecell{Polarization and amplitude\\ beam splitter} & \makecell{Limited bandwidth}\\
\hline
~\cite{meng2020optical} & \makecell {MDM GSP gradient \\ ($\mathrm{Au}$/$\mathrm{SiO_{2}}$/$\mathrm{Au}$)} & \makecell{GSP and SPP\\ mode excitation} & 1 & $850$ & \makecell{SPP coupling and\\ beam steering} & \makecell{Limited bandwidth}\\
\hline

~\cite{zhai2017multiple} & \makecell {Metallic grating MDM \\ ($\mathrm{Ag}$/$\mathrm{SiO_{2}}$/$\mathrm{Ag}$)} & \makecell{GSP + Fabry–Perot \\ resonances}  & 5 & \makecell{$514$, $561$, $660$, \\$819$, $1090$} & \makecell{Simulated\\ multiband absorption} & \makecell{No experimental \\support}\\
\hline

~\cite{ding2020gap} & \makecell {MDM GSP metasurface \\ ($\mathrm{Au}$/$\mathrm{SiO_{2}}$/$\mathrm{Au}$)} & \makecell{GSP resonances}  & 1 & \makecell{$800$--$900$} & \makecell{Multifunctional GSP-\\based metalenses} & \makecell{Limited bandwidth}\\
\hline

~\cite{cai2021dual} & \makecell {MDM GSP metasurface \\ ($\mathrm{Au}$/$\mathrm{SiO_{2}}$/$\mathrm{Au}$)} & \makecell{GSP resonances}  & 1 & $850$ & \makecell{Dual-functional \\optical waveplate} & \makecell{Limited bandwidth}\\
\hline

~\cite{damgaard2020demonstration} & \makecell {Detuned GSP resonator \\ ($\mathrm{Au}$/$\mathrm{SiO_{2}}$/$\mathrm{Au}$)} & \makecell{GSP resonances}  & 2 & \makecell{$785$, $1075$} & \makecell{Optical beam \\ steering} & \makecell{Limited bandwidth}\\
\hline

\textbf{\makecell{This\\work} } & \makecell {Hybrid GSP metasurface \\ ($\mathrm{Al}$/$\mathrm{SiO_{2}}$/$\mathrm{Al}$)} & \makecell {GSP + LSP\\ resonances} & 4 & \makecell{$907$, $1016$, $1111$, \\ $1494$} & \makecell{Multiband, \\ large bandwidth,\\ strong SHG} & $-$ \\
\end{tabular}
\end{sidewaystable}


 



\vfill

\end{document}